\documentclass[11pt,a4paper]{article}
\usepackage{authblk}
\usepackage{lmodern}
\usepackage{jheppub}
\usepackage{amsmath}
\usepackage{mathtools}
\usepackage{tikz}
\usepackage{empheq}
\usepackage{enumitem}
\usepackage{graphics}
\usepackage{wrapfig}
\usepackage{caption}
\usepackage{natbib}
\usepackage{xcolor}
\usepackage{framed}
\usepackage{array}
\usepackage{dsfont}
\definecolor{shadecolor}{gray}{0.925}

\usepackage[cbgreek]{textgreek}

\def\sideremark#1{\ifvmode\leavevmode\fi\vadjust{\vbox to0pt{\vss
 \hbox to 0pt{\hskip\hsize\hskip1em
 \vbox{\hsize3cm\tiny\raggedright\pretolerance10000
 \noindent #1\hfill}\hss}\vbox to8pt{\vfil}\vss}}}%

\newcommand{\bi}{\begin{itemize}}
\newcommand{\ei}{\end{itemize}}
\newcommand{\bea}{\begin{align}}
\newcommand{\eea}{\end{align}}
\newcommand{\be}{\begin{equation}}
\newcommand{\ee}{\end{equation}}

\makeatletter
\renewcommand*\env@matrix[1][\arraystretch]{%
  \edef\arraystretch{#1}%
  \hskip -\arraycolsep
  \let\@ifnextchar\new@ifnextchar
  \array{*\c@MaxMatrixCols c}}
\makeatother
\author{Francesca PACIFICO}
\author{\quad Paolo PERGOLA}
\author{\quad Charlotte SLEIGHT}

\affiliation{Dipartimento di Fisica ``Ettore Pancini'', Universit\`a degli Studi di Napoli Federico II, \\Monte Sant'Angelo, Via Cintia, 80126 Napoli, Italy}

\affiliation{INFN, Sezione di Napoli, Monte Sant'Angelo, Via Cintia, 80126 Napoli, Italy}

\emailAdd{francesca.pacifico@unina.it, paolo.pergola@unina.it, charlotte.sleight@na.infn.it}

\title{\centering \Huge Celestial Mellin Amplitudes}

\abstract{Holographic correlators on the celestial sphere of Minkowski space were recently defined in \cite{Sleight:2023ojm} as the extrapolation of bulk time-ordered correlation functions to the celestial sphere. In this work we explore the Mellin representation of such celestial correlators, which is based on the Mellin representation of conformal correlators introduced by Mack in 2009. Perturbative celestial Mellin amplitudes have some similarities with Mellin amplitudes for AdS Witten diagrams: They are meromorphic functions of the Mellin variables, where contact diagrams have polynomial Mellin amplitudes and particle exchanges are encoded by a specific set of poles. We find the Mellin representation to be a useful tool to study and to compute celestial correlators, and we give various examples in scalar field theories at both tree and loop level, for both massive and massless fields. We also discuss the non-perturbative structure of celestial Mellin amplitudes following from the K\"all\'en-Lehmann spectral representation of bulk two-point functions.}

\begin{document}

\begin{flushright}    
\texttt{}
\end{flushright}

\maketitle

\newpage

\section{Introduction}

The celestial holography programme \cite{Raclariu:2021zjz,Pasterski:2021rjz,McLoughlin:2022ljp,Pasterski:2021raf} seeks to realise the holographic principle in Minkowski space with observables residing on a co-dimension two celestial sphere at null infinity. Lorentz transformations of the bulk Minkowski space act on the celestial sphere like conformal transformations and accordingly the observables on the celestial sphere are those of a conformal field theory (CFT). Such ``Celestial CFTs" however are not garden variety Euclidean Conformal Field Theories since they are not defined by the standard Wick rotation from Minkowski space and therefore are not subject to the usual Wightman \cite{Streater:1989vi} or Osterwalder-Schrader axioms \cite{Osterwalder:1973dx,Osterwalder:1974tc} or, from a more modern perspective, the Euclidean axioms of the Conformal Bootstrap \cite{Kravchuk:2021kwe}. 

\vskip 4pt
The basic observable in quantum gravity in Minkowski space is the S-matrix. This naturally gives rise to a definition of observables on the celestial sphere as a certain change of basis of scattering amplitudes that makes the conformal symmetry manifest \cite{Pasterski:2016qvg,Pasterski:2017kqt}, which we refer to as \emph{celestial amplitudes}. This has led to many insights into the structure and symmetries of the corresponding celestial CFTs by translating knowledge from S-matrix theory. Another approach to celestial holography has been to try to draw lessons from the AdS-CFT correspondence.\footnote{For celestial amplitudes \cite{Pasterski:2016qvg,Pasterski:2017kqt}, various works have aimed to draw lessons from the AdS-CFT correspondence e.g. through the hyperbolic foliation \eqref{HYPFOL} of Minkowski space \cite{deBoer:2003vf,Cheung:2016iub,Casali:2022fro,Iacobacci:2022yjo,Melton:2023bjw,Melton:2024jyq} and the flat limit \cite{deGioia:2022fcn,deGioia:2023cbd}.} In the absence of a standard notion of an S-matrix in anti-de Sitter (AdS) space, holographic correlators can be defined as the extrapolation of bulk correlation functions to the time-like boundary at infinity. From this perspective, one might instead define holographic observables on the celestial sphere of Minkowski space as the extrapolation of bulk correlation functions to the celestial sphere \cite{Sleight:2023ojm}, which we refer to as \emph{celestial correlation functions} \eqref{ccdefn}. These take (off-shell) correlation functions as the fundamental objects, while for celestial amplitudes the fundamental objects are (on-shell) S-matrix elements. Similar to AdS space, massive particles to not reach the boundary at infinity in finite time. However, being off-shell, defining holographic correlators as the extrapolation of bulk correlation functions to the boundary at infinity places massive and massless fields on a similar footing.

\vskip 4pt
The extrapolation of bulk correlation functions to the celestial sphere naturally arises from the hyperbolic foliation of Minkowski space. Analogous to spherical coordinates, in the hyperbolic foliation one writes $X=R{\hat X}$ with radial coordinate $R>0$ and ${\hat X}$ parametrizes the extended unit hyperboloid \cite{deBoer:2003vf}:
\begin{align}\label{HYPFOL}
{\hat X}^2 = \begin{cases}
    +1& \text{for} \quad X^2>0,\\
    -1& \text{for} \quad X^2<0.
\end{cases}
\end{align}
The celestial sphere is then identified with the projective null cone of light rays:
\begin{equation}
Q^2=0, \qquad Q \equiv \lambda Q, \qquad \lambda \in \mathbb{R}^+,
\end{equation}
which is the common boundary of each hyperboloid in the foliation. The extrapolation of a bulk (time-ordered) correlation function of fields $\phi_i\left(X_i\right)$ to the celestial sphere is then implemented by taking a Mellin transform in the radial direction followed by the boundary limit in the hyperbolic directions:
\begin{align}\label{ccdefn}
    \left\langle O_{1}(Q_1)\ldots O_{n}(Q_n)\right\rangle =\prod_i \lim_{{\hat X}_i\to Q_i}\,\int^\infty_0 \frac{{\rm d}R_i}{R_i}\,R_i^{\Delta_i}\left\langle\phi_1(R_1\hat{X}_1)\ldots \phi_n(R_n\hat{X}_n)\right\rangle.
\end{align}  
The Feynman rules for such celestial correlation functions then follow from the Feynman rules for position space time-ordered correlation functions in Minkowski space and were detailed in \cite{Iacobacci:2024nhw}. Evaluating celestial correlators \eqref{ccdefn} requires effective position-space techniques for Minkowski correlation functions, which are relatively scarce compared to their momentum space counterparts. In this work, we develop a perturbative approach to position space correlation functions based on the Schwinger parametrization of the Feynman propagator. This reduces integrals over internal points to Gaussian integrals and naturally gives rise to a Mellin representation of celestial correlation functions \eqref{ccdefn}, which we describe in the following.

\vskip 4pt
Celestial correlators \eqref{ccdefn} share some similarities to the structure to boundary correlators in AdS space and in fact, they can be perturbatively re-written in terms of Witten diagrams in Euclidean anti-de Sitter space (EAdS) \cite{Iacobacci:2024nhw}. This suggests that we might import techniques and results from AdS-CFT to celestial CFT. An important tool that has led to several key insights in the study of boundary correlation functions in AdS is the Mellin representation of conformal correlation functions. Proposed by Mack in his seminal work \cite{Mack:2009mi,Mack:2009gy}, the Mellin representation of a conformal correlation function of operators $O_i$ is given by 
\begin{equation}\label{MR}
\left\langle O_{1}(Q_1)\ldots O_{n}(Q_n)\right\rangle = \int^{i\infty}_{-i\infty}  \frac{{\rm d}\delta_{ij}}{2\pi i} M\left(\delta_{ij}\right)\prod_{i<j} \Gamma\left(\delta_{ij}\right)\left(-2\, Q_i \cdot Q_j\right)^{-\delta_{ij}},
\end{equation}
where $M\left(\delta_{ij}\right)$ is the \emph{Mellin amplitude} and the Mellin integration variables obey the constraints:
\begin{equation}\label{MC}
    \sum^n_{j=1} \delta_{ij}=0, \quad  \delta_{ij} =  \delta_{ji},  \qquad \delta_{ii}=-\Delta_i.
\end{equation}
Mack realised that there is a strong similarity between Mellin amplitudes $M\left(\delta_{ij}\right)$ and flat-space scattering amplitudes. In this regard, it is instructive to introduce the following analogues of the Mandelstam invariants \cite{Costa:2012cb}:
\begin{subequations}\label{MBmandel}
\begin{eqnarray}
     s_{12} = \Delta_1+\Delta_2-2\delta_{12},\\
     s_{13} = \Delta_3-\Delta_4-2\delta_{13}.
\end{eqnarray}
\end{subequations}
For example, in terms of the $s_{ij}$ the Mellin representation of conformal four-point functions takes the form
\begin{multline}\label{MBst}
\hspace*{-0.75cm}\left\langle O_{1}(Q_1)\ldots O_{4}(Q_4)\right\rangle = \frac{1}{\left(-2 Q_1 \cdot Q_2\right)^{\frac{\Delta_1+\Delta_2}{2}}\left(-2 Q_3 \cdot Q_4\right)^{\frac{\Delta_3+\Delta_4}{2}}}\left(\frac{Q_2 \cdot Q_4}{Q_1 \cdot Q_4}\right)^{\frac{\Delta_1-\Delta_2}{2}}\left(\frac{Q_1 \cdot Q_4}{Q_1 \cdot Q_3}\right)^{\frac{\Delta_3-\Delta_4}{2}}\\ \times \int^{+i\infty}_{-i\infty}\frac{{\rm d}s_{12}{\rm d}s_{13}}{\left(4\pi i\right)^2}\, u^{\frac{s_{12}}{2}}v^{-\left(\frac{s_{12}+s_{13}}{2}\right)}\,M\left(s_{12},s_{13}\right)\Gamma\left(\frac{\Delta_1+\Delta_2-s_{12}}{2}\right)\Gamma\left(\frac{\Delta_3+\Delta_4-s_{12}}{2}\right)\\ \times \Gamma\left(\frac{\Delta_{34}-s_{13}}{2}\right)\Gamma\left(\frac{-\Delta_{12}-s_{13}}{2}\right)\Gamma\left(\frac{s_{12}+s_{13}}{2}\right)\Gamma\left(\frac{s_{12}+s_{13}+\Delta_{12}-\Delta_{34}}{2}\right),
\end{multline}
where $\Delta_{ij}=\Delta_i-\Delta_j$ and conformal invariant cross ratios
\begin{equation}
    u = \frac{Q_1 \cdot Q_2\,Q_3 \cdot Q_4}{Q_1 \cdot Q_3 Q_2 \cdot Q_4}, \qquad v = \frac{Q_1 \cdot Q_4\,Q_2 \cdot Q_3}{Q_1 \cdot Q_3 Q_2 \cdot Q_4}.
\end{equation}

The Mellin amplitude $M\left(s_{ij}\right)$ is crossing symmetric and meromorphic with simple poles at
\begin{equation}
s_{ij}=\Delta_k-\ell_k+2m, \qquad m=0,1,2,\ldots,
\end{equation}
where $\Delta_k$ and $\ell_k$ are the scaling dimension and spin of an operator $O_k$ present in the operator product expansions $O_1 O_2 \sim  C_{12k} O_k$ and $O_3 O_4 \sim  C_{34k} O_k$ with OPE coefficients $C_{ijk}$.

\vskip 4pt
Building on the work of Mack, in \cite{Penedones:2010ue} it was proposed that the Mellin amplitude $M\left(s_{ij}\right)$ should be regarded as the AdS scattering amplitude. Like for flat space scattering amplitudes, contact interactions in AdS space give rise to Mellin amplitudes that are polynomial in the ``Mandelstam invariants" $s_{ij}$ and the Mellin amplitudes for tree level exchange diagrams all have poles corresponding to operators dual to the fields exchanged in AdS. Perturbative boundary correlators in AdS also contain contributions from composite (``double-trace") operators (and their descendants) of the schematic form
\begin{equation}\label{DTops}
     \left[O_1O_2\right]_{n,\ell}\sim O_1 \partial_{i_1} \ldots  \partial_{i_\ell}  (\partial^ 2)^ n O_2+\ldots, \qquad \left[O_3O_4\right]_{n,\ell}\sim O_3 \partial_{i_1} \ldots  \partial_{i_\ell} (\partial^ 2)^ n O_4+\ldots\,.
\end{equation}
In the Mellin representation, these are generated by the poles 
\begin{align}\label{DTpoles0}
    s_{12} = \Delta_1+\Delta_2+2n, \qquad s_{12} = \Delta_3+\Delta_4+2n, \qquad n=0,1,2, \ldots,
\end{align}
which are encoded in the $\Gamma$-functions in the integrand \eqref{MBst}.

\vskip 4pt
Inspired by the utility of Mellin amplitudes for holographic correlators in AdS-CFT, in this work we introduce the Mellin representation \eqref{MR} of celestial correlators \eqref{ccdefn}, referring to the corresponding Mellin amplitudes $M\left(\delta_{ij}\right)$ as \emph{celestial Mellin amplitudes}.\footnote{For celestial Mellin amplitudes \eqref{MR}, since the $Q_i$ can be null separated there is an $i\epsilon$ prescription $Q_{i} \cdot Q_j \to Q_{i} \cdot Q_j-i\epsilon$. See appendix D of \cite{Sleight:2023ojm}. This $i\epsilon$ prescription is inherited from that of the Feynman propagator.} Focusing on theories of scalar fields,\footnote{The extension to fields with spin will be presented in \cite{toappear}.} we find they have the following properties in perturbation theory:
\begin{itemize}
    \item {\bf Meromorphicity.} Celestial Mellin amplitudes are meromorphic functions of the Mellin variables $\delta_{ij}$. 
    \item {\bf Contact interactions.} Give rise to polynomial celestial Mellin amplitudes.
    \item {\bf Particle exchanges.} Exchanged particles in Minkowski space are encoded in celestial Mellin amplitudes by infinite families of poles. For four-point correlators \eqref{MBst} with massless external scalar fields, these are the following two families of double-trace-like poles ($n=0,1,2, \ldots$):
\begin{align}\label{Iexchpoles}
   s_{12} = \Delta_1+\Delta_2+(2-d)+2n, \quad s_{12} = \Delta_3+\Delta_4+(2-d)+2n.
\end{align}
For exchanges between massive external scalar fields the pole structure encoding particle exchanges is more involved and is discussed in appendix \ref{A::MassiveExt}.

\vskip 4pt
 When the exchanged particle is also massless, the double-trace poles \eqref{DTpoles0} are absent for integer $d \geq 3$ since the corresponding $\Gamma$-functions in the Mellin integrand \eqref{MBst} are canceled by the Mellin amplitude.
\end{itemize}

\vskip 4pt
In section \ref{sec::MA4pcc} we explain how celestial Mellin amplitudes can be computed in perturbation theory by evaluating position space time-ordered Minkowski correlation functions using the Schwinger parametrization of the Feynman propagator. These techniques extend the approach of \cite{Penedones:2010ue,Paulos:2011ie} which employed the Schwinger parametrization AdS propagators in embedding space to obtain Mellin amplitudes for AdS Witten diagrams. In section \ref{sec::nonpertMA} we explain that these techniques can also be used to study celestial correlators non-perturbatively via the K\"all\'en-Lehmann spectral representation.

\begin{figure}[t]
    \centering
    \includegraphics[width=0.3\textwidth]{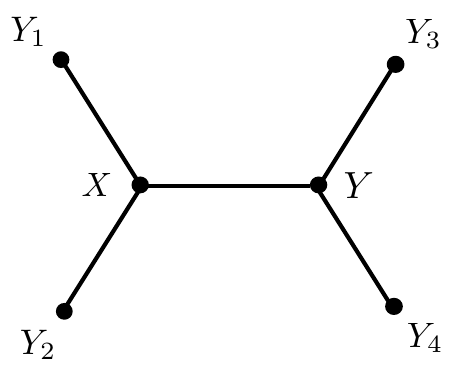}
    \caption{Four-point exchange diagram generated by cubic vertices \eqref{cubicv}.}
    \label{fig::exch}
\end{figure}

\vskip 4pt
We obtain new, closed form, results for celestial correlation functions \eqref{ccdefn} both at tree and loop level, and both for massive and massless fields. For example, the four-point tree-level exchange diagram of a particle of mass $m$ between massless external scalar fields (figure \ref{fig::exch}) with non-derivative cubic couplings has Mellin amplitude (section \ref{subsec::exch}): 
\begin{shaded}
\noindent \emph{Tree-level exchange of a massive particle (mass $m$) between massless external scalars.}
\begin{multline}
    M^{{\cal V}_{12 m}{\cal V}_{34 m}}_{\Delta_1 \Delta_2\Delta_3\Delta_4}\left(s_{12},s_{13}\right)=g_{12}g_{34}\,i^{-\frac{1}{2}\sum_i \Delta_i}\pi^{d+2}\prod^4_{i=1}\frac{1}{4 \pi^{\frac{d+2}{2}}}\Gamma\left(\tfrac{d}{2}-\Delta_i\right)\,\\ \times \frac{1}{4\pi^{\frac{d+2}{2}}} \left(\frac{m}{2}\right)^{3d-4-\sum_i\Delta_i}\Gamma\left(\tfrac{-3d+4+\sum_i\Delta_i}{2}\right) \frac{\Gamma\left(\tfrac{2-d+\Delta_1+\Delta_2-s_{12}}{2}\right)\Gamma\left(\tfrac{2-d+\Delta_3+\Delta_4-s_{12}}{2}\right)}{\Gamma\left(\tfrac{d+2-\Delta_1-\Delta_2-s_{12}}{2}\right)\Gamma\left(\tfrac{d+2-\Delta_3-\Delta_4-s_{12}}{2}\right)}\,\\ \times {}_3F_2\left(\begin{matrix}\frac{d+2- \Delta_1- \Delta_2- \Delta_3- \Delta_4}{2},\tfrac{2-d+\Delta_1+\Delta_2-s_{12}}{2},\tfrac{2-d+\Delta_3+\Delta_4-s_{12}}{2}\\\tfrac{d+2-\Delta_3-\Delta_4-s_{12}}{2},\tfrac{d+2-\Delta_1-\Delta_2-s_{12}}{2}\end{matrix};1\right).\label{introtlexch}
    \end{multline}
\end{shaded}
\noindent These results straightforwardly extend to exchange diagrams at loop level and beyond, which have the same dependence on the ``Mandelstam invariants" $s_{12}$ and $s_{13}$. Using the K\"all\'en-Lehmann spectral representation of the exact two-point function, the Mellin amplitude for such diagrams (see figure \ref{fig::4ptexch}) takes the form (section \ref{sec::nonpertMA}):
\begin{shaded}
\noindent \emph{General four-point exchange between massless external scalars.}
\begin{equation}
    M^{{\cal V}_{12 \Gamma}{\cal V}_{34 \Gamma}}_{\Delta_1 \Delta_2\Delta_3\Delta_4}\left(s_{12},s_{13}\right)= {\cal M}\left[\rho\right]\left(\tfrac{3d-2-\sum_i\Delta_i}{2}\right)M^{{\cal V}_{12 m=1}{\cal V}_{34 m=1}}_{\Delta_1 \Delta_2\Delta_3\Delta_4}\left(s_{12},s_{13}\right),
    \end{equation}
\end{shaded}
\noindent where ${\cal M}\left[\rho\right]\left(\bullet\right)$ is the Mellin transform of the spectral function $\rho(\mu^2)$ for the interacting bulk two-point function $\Gamma\left(X,Y\right)$ and encodes the dependence on the masses of exchanged particles.

\vskip 4pt
We also study the massless limit of perturbative celestial correlators \eqref{ccdefn}, which for scalar fields is smooth and matches with expressions that one would obtain using the Feynman propagator for massless scalar fields directly. This shows that massless limit commutes with the prescription \eqref{ccdefn} to extrapolate bulk time-ordered correlation functions to the celestial sphere. Mellin amplitudes involving massless scalars are simpler than their massive counterparts owing to the simplicity of the Feynman propagator in this case, which is a power law. When all fields in the diagram are massless, the latter implies linear constraints on the scaling dimensions. For example, for integer dimensions $d \ge 3$ the $m \to 0$ limit of the exchange diagram \eqref{introtlexch} is (section \ref{subsec::exch} equation \eqref{mldge3}):
\begin{shaded}
\noindent \emph{Tree-level exchange for massless scalars for integer $d \geq 3$.}
\begin{multline}
    M^{{\cal V}_{12 \varphi}{\cal V}_{34 \varphi}}_{\Delta_1 \Delta_2\Delta_3\Delta_4}\left(s_{12},s_{13}\right)=g_{12}g_{34}\,\pi^{d+2}\,\prod^4_{i=1}\frac{1}{4\pi^{\frac{d+2}{2}}}\Gamma\left(\tfrac{d}{2}-\Delta_i\right)\,\\ \times (-1)^{3-d} \frac{\Gamma\left(\frac{d}{2}-1\right)}{\Gamma\left(2-\frac{d}{2}\right)} \frac{\Gamma\left(2d-3-\Delta_1-\Delta_2\right)}{\Gamma\left(d-\Delta_1-\Delta_2\right)} \\ \times \frac{1}{4\pi^{\frac{d+2}{2}}}2\pi i\, \delta\left(\frac{-3d+4+\sum_i\Delta_i}{2}\right) \frac{\Gamma \left(\frac{2-d+\Delta_3+\Delta_4-s_{12}}{2}\right) \Gamma \left(\frac{2-d+\Delta_1+\Delta_2-s_{12}}{2}\right)}{\Gamma\left(\frac{\Delta_3+\Delta_4-s_{12}}{2}\right)\Gamma\left(\frac{\Delta_1+\Delta_2-s_{12}}{2}\right)},
\end{multline}
\end{shaded}
\noindent where the constraint on the scaling dimensions $\Delta_i$ simplifies the hypergeometric function ${}_3F_2$ to a simple ratio of $\Gamma$-functions for integer $d \geq 3$. The latter is in fact proportional to the Mellin amplitude representation of a single conformal partial wave ${\cal F}_{\Delta,0}$ (see e.g. \cite{Sleight:2018epi} equation (3.6)) encoding the contributions from two exchanged (shadow) conformal multiplets of scaling dimension:
\begin{equation}\label{mlrepsint}
    \Delta=\Delta_{1}+\Delta_2+2-d, \qquad d-\Delta=\Delta_{3}+\Delta_4+2-d.
\end{equation}
These are shadow of one another by virtue of the delta function constraint on the scaling dimensions in massless correlators. The exchange diagram in massless scalar theories therefore has a simple explicit form given by a (single-valued) sum of two conformal blocks corresponding to the representations \eqref{mlrepsint}. For $d=2$ the expression for the exchange also has contributions from the derivatives of these conformal blocks and the Mellin amplitude is given in equation \eqref{mlexchd2}. These features moreover extend to exchanges of massless fields with non-trivial spin, which will be presented in the upcoming work \cite{toappear}.

\vskip 4pt
The outline of the paper is as follows: In section \ref{sec::MA4pcc} we outline the general approach to compute perturbative time-ordered correlators in position space using Schwinger parametrization of the Feynman propagator and how to extract the Mellin amplitude of the corresponding correlator on the celestial sphere. In section \ref{subsec::contact} we consider the celestial Mellin amplitudes generated by contact diagrams generated by derivative and non-derivative interactions in scalar field theories. In section \ref{subsec::exch} we calculate the celestial Mellin amplitude generated by a four-point tree-level exchange diagram with external massless scalars, for both massive and massless exchanged scalars. The extension to massive external scalars is given in appendix \ref{A::MassiveExt}. In section \ref{subsec::1loop} we consider the celestial Mellin amplitudes generated by four-point ``Melonic" loop diagrams, which are formed by adding lines between a pair of internal points. In section \ref{sec::nonpertMA} we discuss the non-perturbative structure of celestial Mellin amplitudes using the K\"all\'en-Lehmann spectral representation. In appendix \ref{A::DI} further details are given on the evaluation of various integrals encountered in this work.

\newpage

\section{Mellin amplitudes for perturbative celestial correlators}
\label{sec::MA4pcc}

In this section we explain how to determine the Mellin amplitudes \eqref{MR} for celestial correlators \eqref{ccdefn} in perturbation theory by evaluating Minkowski time-ordered correlators using the Schwinger parametrization of the Feynman propagator. For ease of illustration we will focus on theories of scalar fields. 

\vskip 4pt
A convenient approach to position space correlation functions in Minkowski space is based on the Mellin transform of the exponential function and its inverse:
\begin{align}\label{schwp}
    z^{-\delta}&=\frac{1}{\Gamma\left(\delta\right)} \int^\infty_0 \frac{{\rm d}t}{t} t^\delta\,e^{-tz},\\ \label{expMB}
    e^{-z} &= \int^{+i \infty}_{-i\infty} \frac{{\rm d} \delta}{2\pi i} \Gamma\left(\delta\right) z^{-\delta}.
\end{align}
The position space Feynman propagator for a scalar field of mass $m$ in $\mathbb{M}^{d+2}$ is
\begin{align}
    G^{(m)}_{T}\left(X,Y\right)&=\frac{1}{2 \pi^{\frac{d+2}{2}}}\left(\frac{m}{2}\right)^{\frac{d}{2}} \frac{1}{\left[\left(X-Y\right)^2+i\epsilon\right]^{\frac{d}{4}}}K_{\frac{d}{2}}\left(m\left[\left(X-Y\right)^2+i\epsilon\right]^{\frac{1}{2}}\right),\\
    &=\int^{+i\infty}_{-i\infty}\frac{{\rm d}s}{2\pi i}\frac{1}{4 \pi^{\frac{d+2}{2}}}\Gamma\left(s+\tfrac{d}{4}\right)\Gamma\left(s-\tfrac{d}{4}\right)\left(\frac{m}{2}\right)^{-2s+\frac{d}{2}}\\
    & \hspace*{7cm} \times \left[\left(X-Y\right)^2+i\epsilon\right]^{-\left(s+\frac{d}{4}\right)}, \nonumber
\end{align}
where on the second line we inserted the Mellin-Barnes representation of the Bessel-$K$ function. Using the Mellin transform \eqref{schwp}, the dependence on the separation $\left(X-Y\right)^2$ can be exponentiated (i.e. Schwinger parametrization):
\begin{equation}\label{SPXY}
    \left[\left(X-Y\right)^2+i\epsilon\right]^{-\alpha}= \frac{i^{-\alpha}}{\Gamma\left(\alpha\right)} \int^{\infty}_0 \frac{{\rm d}t}{t} t^{\alpha} \exp \left[i t \left(X-Y\right)^2\right].
\end{equation}
For massless scalars $m=0$, this is simply
\begin{align}\label{Spml}
    G^{(0)}_{T}\left(X,Y\right)&=\frac{\Gamma\left(\frac{d}{2}\right)}{4 \pi^{\frac{d+2}{2}}} \frac{1}{\left[\left(X-Y\right)^2+i\epsilon\right]^{\frac{d}{2}}},\\
    &=\frac{i^{-\frac{d}{2}}}{4 \pi^{\frac{d+2}{2}}} \int^{\infty}_0 \frac{{\rm d}t}{t} t^{\frac{d}{2}} \exp \left[i t \left(X-Y\right)^2\right],\label{MLschw}
\end{align}
while for scalars with generic mass there is an additional Mellin-Barnes integral in the mass: 
\begin{multline}\label{SPm}
    G^{(m)}_{T}\left(X,Y\right)= \int^{+i\infty}_{-i\infty}\frac{{\rm d}s}{2\pi i}\frac{1}{4\pi^{\frac{d+2}{2}}}\Gamma\left(s-\tfrac{d}{4}\right)\left(\frac{m}{2}\right)^{-2s+\frac{d}{2}}i^{-(s+\tfrac{d}{4})}\\
\times \int^{\infty}_0 \frac{{\rm d}t}{t} t^{s+\tfrac{d}{4}}\, \exp \left[i t \left(X-Y\right)^2\right].
\end{multline}
The Schwinger parametrization \eqref{SPXY} thus reduces integrals over internal points in Feynman diagrams to (nested) Gaussian integrals of the form (see appendix \ref{A::GI})
\begin{multline}
    \int {\rm d}^{d+2}X \exp\left[i \sum\limits_i t_i\left(X-Y_i\right)^2\right]\\ = i^{\frac{d}{2}}\left(\frac{\pi}{t_1 + \ldots +t_n}
    \right)^{\frac{d+2}{2}} \exp\left[-\frac{i}{t_1+ \ldots t_n}\left(\sum_{i}t_i Y_i\right)^2+i\sum_i t_i Y^2_i\right].
\end{multline}

\vskip 4pt
To obtain the corresponding celestial correlation function \eqref{ccdefn}, we take the Mellin transform of the external points $Y_i=R_i {\hat Y}_i$ with respect to the $R_i$ and send ${\hat Y}_i \to Q_i$. In this limit, since $Q^2_i=0$ the $Y^2_i$ terms vanish and one is left with a function of the $Q_i \cdot Q_j$. The Mellin representation \eqref{MR} then follows from the inverse Mellin transform \eqref{expMB}:
\begin{equation}
  \hspace*{-0.5cm}  \exp \left[- i \sum\limits_{i<j}c_{ij} R_iR_j 2 Q_i \cdot Q_j\right] = \int^{i\infty}_{-i\infty}  \prod_{i<j}\,\frac{{\rm d}\delta_{ij}}{2\pi i}\, \left(ic_{ij} R_i R_j \left(2Q_i \cdot Q_j-i\epsilon\right)\right)^{-\delta_{ij}} \Gamma\left(\delta_{ij}\right),
\end{equation}
for some $c_{ij}$ which depend on the Schwinger parameters of the Feynman propagators. The constraints \eqref{MC} on the Mellin variables (from dilatation invariance) arise from the Mellin transform with respect to the radial coordinates $R_i$:
    \begin{equation}
    \int_0^\infty\frac{{\rm d}R_i}{R_i} R^{\Delta_i}_iR_i^{-\sum_{j\neq i}\delta_{ij}}=(2\pi i)\delta\left(\Delta_i-\sum_{j\neq i}\delta_{ij}\right).
\end{equation}
Determining the celestial Mellin amplitude for a given process in Minkowski space is then reduced to evaluating the integrals over the Schwinger parameters for each Feynman propagator. The Schwinger-parameter integrals encountered for the diagrams considered in this work are detailed in appendix \ref{A::SI}.

\vskip 4pt
These techniques extend to celestial correlators \eqref{ccdefn} the approach of \cite{Penedones:2010ue,Paulos:2011ie} which was used to obtain Mellin amplitudes for AdS Witten diagrams. In the remainder of this section we use the above techniques to determine the Mellin amplitudes for various types of diagrams in scalar field theories in $(d+2)$ dimensional Minkowski space $\mathbb{M}^{d+2}$, recovering previous results \cite{Sleight:2023ojm} for contact diagrams and obtaining new ones for particle exchanges and loops. These will illustrate the simplicity of the representation and give some intuition for their general structure in perturbation theory. 

\paragraph{Notation.} To illustrate the approach we will focus on diagrams with no more than two internal points. The Schwinger parameters for external legs are denoted by $t_i$ and those for internal legs are denoted by ${\bar t}_i$. For propagators corresponding to massive particles \eqref{SPm}, the Mellin variables for external and internal legs are denoted by $s_i$ and ${\bar s}_i$ respectively.

\subsection{Contact diagrams}
\label{subsec::contact}

Consider the following non-derivative interaction of $n$ scalar fields $\phi_i$ of generic mass $m_i$, $i=1, \ldots, n$, 
\begin{equation}\label{vnptcont}
    {\cal V}_{12 \ldots n} = g_{12 \ldots n}\, \phi_1 \phi_2 \ldots \phi_n.
\end{equation}
The corresponding contact diagram contribution to the time-ordered $n$-point function of $\phi_i$ is
\begin{equation}\label{ncontact}
    {\cal A}^{{\cal V}_{12 \ldots n}}\left(Y_1,\ldots,Y_n\right)=-i g \int {\rm d}^{d+2}X\, G^{(m_1)}_{T}\left(X,Y_1\right) \ldots G^{(m_n)}_{T}\left(X,Y_n\right).
\end{equation}
Employing Schwinger parametrization \eqref{SPm} of the Feynman propagators, this reads
\begin{multline}
 \hspace*{-0.5cm}   {\cal A}^{{\cal V}_{12 \ldots n}}\left(Y_1,\ldots,Y_n\right) = - ig_{12 \ldots n}\,  \int^{+i\infty}_{-i\infty}\prod^n_{i=1}\frac{{\rm d}s_i}{2\pi i}\frac{1}{4 \pi^{\frac{d+2}{2}}}\Gamma\left(s_i-\tfrac{d}{4}\right)\left(\frac{m_i}{2}\right)^{-2s_i+\frac{d}{2}} i^{-(s_i+\frac{d}{4})}\\ \times \int^{\infty}_0 \prod^n_{i=1} \frac{{\rm d}t_i}{t_i} t^{s_i+\frac{d}{4}}_i\, \int d^{d+2}X \exp\left[i \sum\limits_i t_i\left(X-Y_i\right)^2\right],
\end{multline}
where the integrals over $s_i$ encode the mass dependence of each field. The integral over the internal point is reduced to a Gaussian (see appendix \ref{A::GI}):
\begin{multline}\label{GaussCont}
    \int {\rm d}^{d+2}X \exp\left[i \sum\limits_i t_i\left(X-Y_i\right)^2\right] = i^{\frac{d}{2}}\left(\frac{\pi}{t_1 + \ldots +t_n}
    \right)^{\frac{d+2}{2}} \\ \times \exp\left[-i \left(t_1+ \ldots t_n\right)\left(\frac{t_1 Y_1 + \ldots + t_n Y_n}{t_1+ \ldots t_n}\right)^2+i\sum_i t_iY_i^2\right].
\end{multline}

\begin{figure}[t]
    \centering
    \includegraphics[width=0.3\textwidth]{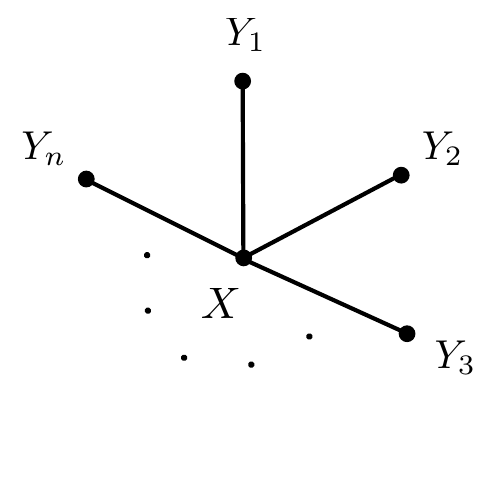}
    \caption{$n$-point contact diagram generated by the vertex \eqref{vnptcont}.}
    \label{fig::cont_npt}
\end{figure}

The contribution from the contact diagram \eqref{ncontact} to the corresponding celestial correlation function is then obtained by plugging in the above result into the definition \eqref{ccdefn} of celestial correlators:
\begin{align}
    {\cal A}^{{\cal V}_{12 \ldots n}}_{\Delta_1 \ldots \Delta_n}\left(Q_1,\ldots,Q_n\right)&=\lim_{{\hat Y}_i \to Q_i}   \int^{\infty}_0 \left(\prod\limits^n_{i=1} \frac{{\rm d}R_i}{R_i} R^{\Delta_i}_i\right) {\cal A}^{{\cal V}_{12 \ldots n}}\left(R_1 {\hat Y}_1,\ldots,R_n {\hat Y}_n\right)\\ \nonumber
    &= - ig_{12 \ldots n} \, i^{\frac{d}{2}} \pi^{\frac{d+2}{2}} \\ \nonumber &\times \int^{\infty}_0\left(\prod\limits^n_{i=1} \frac{{\rm d}R_i}{R_i} R^{\Delta_i}_i\right) \int^{+i\infty}_{-i\infty}\frac{{\rm d}s_i}{2\pi i}\prod^n_{i=1}\frac{1}{4 \pi^{\frac{d+2}{2}}}\Gamma\left(s_i-\tfrac{d}{4}\right)\left(\frac{m_i}{2}\right)^{-2s_i+\frac{d}{2}} i^{-(s_i+\frac{d}{4})}\\ & \hspace*{-1.5cm}   \times \int^{\infty}_0 \prod^n_{i=1} \frac{{\rm d}t_i}{t_i} t^{s_i+\frac{d}{4}}_i\, \left(t_1 + \ldots +t_n\right)^{-\frac{d+2}{2}} \exp\left[-i \frac{2}{t_1+ \ldots t_n}\sum_{i<j} t_i t_j R_i R_j Q_i \cdot Q_j \right], \nonumber
\end{align}
where we use that the $Y^2_i$ terms vanish in the limit ${\hat Y}_i \to Q_i$, since $Q^2_i = 0$. 

\vskip 4pt
The Mellin representation \eqref{MR} of the celestial correlator is then obtained by employing the Mellin-Barnes representation \eqref{expMB} of the exponential function
\begin{multline}
    \exp\left[-i \frac{2}{t_1+ \ldots t_n}\sum_{i<j} t_i t_j R_i R_j Q_i \cdot Q_j \right] \\ = \int^{i\infty}_{-\infty}  \prod_{i<j}\,\frac{{\rm d}\delta_{ij}}{2\pi i}\, \left(\frac{2i t_{i}t_j R_i R_j Q_i \cdot Q_j}{t_1 + \ldots + t_n}\right)^{-\delta_{ij}} \Gamma\left(\delta_{ij}\right),
\end{multline}
which gives the Mellin amplitude
\begin{multline}\label{mellinAmplCont}
    M^{{\cal V}_{12 \ldots n}}_{\Delta_1 \ldots \Delta_n}\left(\delta_{ij}\right)= - ig_{12 \ldots n} \, i^{\frac{d}{2}} \pi^{\frac{d+2}{2}}   \\ \times  \int^{+i\infty}_{-i\infty}\frac{{\rm d}s_i}{2\pi i}\prod^n_{i=1}\frac{1}{4 \pi^{\frac{d+2}{2}}}\Gamma\left(s_i-\tfrac{d}{4}\right)\left(\frac{m_i}{2}\right)^{-2s_i+\frac{d}{2}} i^{-\left(s_i+\frac{1}{2}\left(\frac{d}{2}-\Delta_i\right)\right)}\\
    \times T_{\text{cont.}}\left(s_1+\frac{d}{4}-\Delta_1,\ldots ,s_n+\frac{d}{4}-\Delta_n;\frac{\beta}{2}\right),
\end{multline}
where for convenience we defined
\begin{equation}\label{beta}
    \beta = -(d+2)+\sum_i \Delta_i.
\end{equation}
For contact diagrams the Schwinger integrals take the form (see appendix \ref{A::SI})
\begin{align}\label{schwintc}
    T_{\text{cont.}}\left(a_1,\ldots a_n;b\right)&= \int^{\infty}_0 \prod^n_{i=1} \frac{{\rm d}t_i}{t_i} t^{a_i}_i \left(t_1+\ldots + t_n\right)^{b}= 2 \pi i\, \delta (b+\sum_i a_i) \frac{\prod^n_{i=1}\Gamma\left(a_i\right)}{\Gamma\left(-b\right)}. 
\end{align} 
This gives the Mellin amplitude as 
\begin{multline}\label{MAc}
     M^{{\cal V}_{12 \ldots n}}_{\Delta_1 \ldots \Delta_n}\left(\delta_{ij}\right)=  -g_{12 \ldots n} \,  \pi^{\frac{d+2}{2}}   \\ \times \frac{1}{\Gamma\left(-\frac{\beta}{2}\right)}  \int^{+i\infty}_{-i\infty}\frac{{\rm d}s_i}{2\pi i} \,\left(2\pi i\right)\delta\left(\tfrac{\beta}{2}+\sum_i \left(s_i+\tfrac{1}{2}\left(\tfrac{d}{2}-\Delta_i\right)\right)\right)\\ \times \prod^n_{i=1}\frac{1}{4 \pi^{\frac{d+2}{2}}}\Gamma\left(s_i-\frac{1}{2}\left(\Delta_i - \tfrac{d}{2}\right)\right)\Gamma\left(s_i+\frac{1}{2}\left(\Delta_i - \tfrac{d}{2}\right)\right)\left(\frac{m_i}{2}\right)^{-2s_i+\frac{d}{2}-\Delta_i},
\end{multline}
where we made the change of variable $s_i \to s_i +\frac{\Delta_i}{2}$. This expression matches the result given originally in \cite{Sleight:2023ojm}.\footnote{To see this, combine equations (4.9) and (D.7) of reference \cite{Sleight:2023ojm}. As noted in the latter work, the expression \eqref{MAc} can equivalently be represented as an integral over the radial direction $R$ of a product of Bessel-$K$ functions $K_{\Delta_i-\frac{d}{2}}\left(m_i R\right)$. In Mellin space the radial integral reduces to the Dirac delta function in \eqref{MAc}.} Note that, as for the Mellin amplitudes of non-derivative contact Witten diagrams in AdS, this is independent of the Mellin variables $\delta_{ij}$. At four-points $(n=4)$, the poles in the Mandelstam invariant $s_{12}$ are therefore only at the double-trace values \eqref{DTpoles0}, as in AdS.

\vskip 4pt
Notice that the Mellin amplitude \eqref{MAc} seems to be vanishing for 
\begin{equation}\label{pv}
    \sum_i\Delta_i=(d+2)+2m, \qquad m =0, 1, 2, \ldots,
\end{equation}
i.e. on the poles of the $\Gamma$-function $\Gamma\left(-\beta/2\right)$ in the denominator. This means that when the scaling dimensions satisfy \eqref{pv} the corresponding correlator would be vanishing as well -- at least for boundary points $Q_i$ that are not null separated i.e. $Q_i \cdot Q_j \ne 0$. A similar mechanism occurs for correlation functions on the late-time boundary of dS space \cite{Sleight:2019mgd,Sleight:2019hfp,Bzowski:2023nef}, in which case the corresponding correlator is a contact term (if not zero identically).

\paragraph{Massless scalars.} Diagrams involving massless scalars can be computed in a similar fashion using the Schwinger-parametrization \eqref{Spml} of the massless Feynman propagator. Alternatively, one can take the massless limit of individual legs in the massive contact diagram \eqref{Spml}. In the limit $m \to 0$ we have:
\begin{multline}\label{mllim}
    \Gamma\left(s-\frac{1}{2}\left(\Delta - \tfrac{d}{2}\right)\right)\Gamma\left(s+\frac{1}{2}\left(\Delta - \tfrac{d}{2}\right)\right)\left(\frac{m}{2}\right)^{-2s+\frac{d}{2}-\Delta}\\ \to \Gamma\left(\tfrac{d}{2}-\Delta\right)\,2\pi i\, \delta \left(s+\frac{1}{2}\left(\Delta - \tfrac{d}{2}\right)\right).
\end{multline}
For example, the $n$-point contact diagram for all massless scalars is
\begin{equation}\label{MAcml}
     M^{{\cal V}_{12 \ldots n}}_{\Delta_1 \ldots \Delta_n}\left(\delta_{ij}\right)=  -g_{12 \ldots n} \,  \pi^{\frac{d+2}{2}}  \left(\prod^n_{i=1}\frac{1}{4 \pi^{\frac{d+2}{2}}}\Gamma\left(\tfrac{d}{2}-\Delta_i\right)\right)\frac{1}{\Gamma\left(-\frac{\beta}{2}\right)} \left(2\pi i\right)\delta\left({\tilde \beta}/2\right),
\end{equation}
where 
\begin{equation}
    {\tilde \beta}=-(d+2)+\sum_i {\tilde \Delta}_i,
\end{equation}
which, compared to $\beta$ in \eqref{beta}, is a sum over shadow dimensions ${\tilde \Delta}_i=d-\Delta_i$. The expression \eqref{MAcml} matches with the result that one would obtain using the massless Feynman propagator \eqref{Spml} directly. The massless limit therefore commutes with the extrapolation \eqref{ccdefn} of Minkowski correlators to the celestial sphere.

\vskip 4pt
The Dirac delta function constraint \eqref{MAcml} on the scaling dimensions $\Delta_i$ combined with the zeros \eqref{pv} implies that $n$-point contact diagrams for massless scalars are vanishing (for non-null separated points $Q_i \cdot Q_j \ne 0$) when:
\begin{equation}\label{mlnptvanish}
    (n-2)d = 2m+4, \qquad m=0, 1, 2, \ldots.
\end{equation}
E.g. For three-point contact diagrams $(n=3)$ in even dimensions $d>2$. This will be useful to keep in mind when considering the corresponding exchange diagrams in section \ref{subsec::exch}.

\paragraph{Derivative interactions.} Celestial Mellin amplitudes generated by derivative contact interactions are polynomial in the Mellin variables $\delta_{ij}$. For example, consider the following two-derivative interaction of $n$ scalar fields $\phi_i$ of generic mass $m_i$, $i=1, \ldots, n$:
\begin{equation}
    {\cal V}_{{\bar 1}{\bar 2}3 \ldots n} = g_{12 \ldots n}\, (\partial_\mu\phi_1 \partial^\mu\phi_2)\phi_3 \ldots \phi_n,
\end{equation}
where the notation ${\bar 1}{\bar 2}$ refers to the fields on which the derivatives are acting. By translation invariance,\footnote{Which in particular implies that $\partial^\mu_X G^{(m)}_{T}\left(X,Y\right) = - \partial^\mu_Y G^{(m)}_{T}\left(X,Y\right)$.} the corresponding contribution to the time-ordered $n$-point function of the $\phi_i$ can be written as $\partial_{Y_1}\cdot\partial_{Y_2}$ acting on the non derivative one \eqref{ncontact}:
\begin{equation}
    {\cal A}^{{\cal V}_{{\bar 1}{\bar 2}3 \ldots n}}\left(Y_1,\ldots,Y_n\right)=\left( \partial_{Y_1}\cdot\partial_{Y_2}\right){\cal A}^{{\cal V}_{123 \ldots n}}\left(Y_1,\ldots,Y_n\right).
\end{equation}
The corresponding celestial correlator reads
\begin{multline}\label{DerivInt}
\hspace*{-0.2cm} {\cal A}^{{\cal V}_{{\bar 1}{\bar 2}3 \ldots n}}\left(Q_1,\ldots,Q_n\right)=-ig_{12 \ldots n}\,i^{\frac{d}{2}}\pi^{\frac{d+2}{2}}\int^{+i\infty}_{-i\infty}\prod^n_{i=1}\frac{{\rm d}s_i}{2\pi i}\frac{1}{4\pi^{\frac{d+2}{2}}}\Gamma\left(s_i-\tfrac{d}{4}\right)\left(\frac{m_i}{2}\right)^{-2s_i+\frac{d}{2}} i^{-(s_i+\frac{d}{4})}\\
\times\int^{\infty}_0 \left(\prod\limits^n_{i=1} \frac{{\rm d}R_i}{R_i} R^{\Delta_i}_i\right) \int^{\infty}_0 \prod^n_{i=1} \frac{{\rm d}t_i}{t_i} t^{s_i+\frac{d}{4}}_i\, \left(t_1 + \ldots +t_n\right)^{-\frac{d+2}{2}}\\
\times\left[\frac{-2it_1t_2(d+2)}{t_1+\ldots+t_n}-\frac{8t_1t_2}{(t_1+\ldots+t_n)^2}\sum_{\alpha<\beta}t_\alpha t_\beta R_\alpha R_\beta Q_\alpha\cdot Q_\beta\right.\\
\left.+\frac{1}{t_1+\ldots+t_n}4t_1t_2\left(R_1Q_1+R_2Q_2\right)\cdot(t_1R_1Q_1+\ldots+t_nR_nQ_n)-4t_1t_2R_1R_2Q_1\cdot Q_2\right]\\
 \times\int^{i\infty}_{-i\infty}  \prod_{i<j}\,\frac{{\rm d}\delta_{ij}}{2\pi i}\, \left(\frac{2i t_{i}t_j R_i R_j Q_i \cdot Q_j}{t_1 + \ldots + t_n}\right)^{-\delta_{ij}} \Gamma\left(\delta_{ij}\right),
\end{multline}
where the action of derivatives like $\partial_{Y_1}\cdot\partial_{Y_2}$ generate polynomials in the $Q_i\cdot Q_j$. These, in turn, give rise to polynomials in the Mellin variables $\delta_{ij}$ in the corresponding Mellin amplitude by virtue of the identity:
\begin{multline}
    \int^{+i\infty}_{-i\infty} \frac{{\rm d}\delta_{ij}}{2\pi i} \left( \ldots \right) \left(2i Q_i \cdot Q_j\right)^n  \Gamma\left(\delta_{ij}\right)\left(2i Q_i \cdot Q_j\right)^{-\delta_{ij}} \\ = \int^{+i\infty}_{-i\infty} \frac{{\rm d}\delta_{ij}}{2\pi i} \left( \ldots \right)   \left(\delta_{ij}\right)_n\Gamma\left(\delta_{ij}\right)\left(2i Q_i \cdot Q_j\right)^{-\delta_{ij}},
\end{multline}
where we made the change of variables $\delta_{ij} \to \delta_{ij}+n$ and $\left(z\right)_n$ is the usual Pochhammer symbol, which is a degree $n$ monomial in the variable $z$. Under this change of variables, the integrals over the radial variables $R_i$ are the same as in the non-derivative case, while the integrals \eqref{schwintc} over the Schwinger parameters $t_i$ have shifts in the exponents $a_i$ and $b$:
\begin{multline}
 M^{{\cal V}_{{\bar 1}{\bar 2}3 \ldots n}}_{\Delta_1 \ldots \Delta_n}\left(\delta_{ij}\right)=-2g_{12 \ldots n}\, i^{\frac{d}{2}}\pi^{\frac{d+2}{2}}\\
 \times\int^{+i\infty}_{-i\infty}\prod^n_{i=1}\frac{{\rm d}s_i}{2\pi i}\frac{1}{4\pi^{\frac{d+2}{2}}}\Gamma\left(s_i-\frac{d}{4}\right)\left(\frac{m_i}{2}\right)^{-2s_i+\frac{d}{2}}i^{-(s_i+\frac{d}{4}+\frac{\Delta_i}{2})}\\
\hspace*{-0.75cm} \times\left[\Bigl(d+2-\sum^n_{i=1} \Delta_i\Bigr)T_{\text{cont.}}\left(s_1+\tfrac{d}{4}-\Delta_1+1,s_2+\tfrac{d}{4}-\Delta_2+1,s_3+\tfrac{d}{4}-\Delta_3,\ldots, s_n+\tfrac{d}{4}-\Delta_n;\tfrac{\beta}{2}-1\right)\right.\\
+\Delta_1\, T_{\text{cont.}}\left(s_1+\tfrac{d}{4}-\Delta_1,s_2+\tfrac{d}{4}-\Delta_2+1,s_3+\tfrac{d}{4}-\Delta_3,\ldots, s_n+\tfrac{d}{4}-\Delta_n;\tfrac{\beta}{2}\right)\\+\Delta_2\, T_{\text{cont.}}\left(s_1+\tfrac{d}{4}-\Delta_1+1,s_2+\tfrac{d}{4}-\Delta_2,s_3+\tfrac{d}{4}-\Delta_3,\ldots s_n+\tfrac{d}{4}-\Delta_n;\tfrac{\beta}{2}\right)\\
-\left.\delta_{12}T_{\text{cont.}}\left(s_1+\tfrac{d}{4}-\Delta_1,\ldots, s_n+\tfrac{d}{4}-\Delta_n;\tfrac{\beta}{2}+1\right)\Biggr]\right..
\end{multline}
Inserting the expression for the Schwinger integral \eqref{schwintc} gives the Mellin amplitude:
\begin{multline}\label{MellinDerivInt}
 M^{{\cal V}_{{\bar 1}{\bar 2}3 \ldots n}}_{\Delta_1 \ldots \Delta_n}\left(\delta_{ij}\right)=-2g_{12 \ldots n}\, i^{\frac{d}{2}}\pi^{\frac{d+2}{2}}\frac{1}{\Gamma\left(-\frac{\beta}{2}\right)}\\
 \times\int^{+i\infty}_{-i\infty}\prod^n_{i=1}\frac{{\rm d}s_i}{2\pi i}\frac{1}{4\pi^{\frac{d+2}{2}}}\Gamma\left(s_i-\frac{d}{4}\right)\left(\frac{m_i}{2}\right)^{-2s_i+\frac{d}{2}}i^{-(s_i+\frac{d}{4})}\\
\times(2\pi i)\,\delta\left(\frac{d}{2}-\sum_i \left(s_i+\frac{d}{4}-\frac{\Delta_i}{2}\right)\right)\prod^n_{i=1}\Gamma\left(s_i+\frac{d}{4}-\Delta_i\right)\\
\times\left[2\left(s_1+\frac{d}{4}-\Delta_1\right)\left(s_2+\frac{d}{4}-\Delta_2\right)\right.\\
+\left(\left(s_2+\frac{d}{4}-\Delta_2\right)\Delta_1+\left(s_1+\frac{d}{4}-\Delta_1\right)\Delta_2\right)
-\left.\left(\frac{d}{2}-\frac{1}{2}\sum_{i=1}^n\Delta_i\right)\delta_{12}\right],
\end{multline}
which is linear in the Mellin variable $\delta_{12}$.

\vskip 4pt
From the above example it is straightforward to conclude that a scalar contact diagram in Minkowski space generated by an interaction with 2$N$ derivatives has a Celestial Mellin amplitude that is a degree $N$ polynomial in the Mellin variables $\delta_{ij}$, analogous to Mellin amplitudes \cite{Penedones:2010ue} in anti-de Sitter space.

\subsection{Exchange diagrams}
\label{subsec::exch}

Consider the four-point exchange diagram for a field $\varphi$ of mass $m$ generated by the cubic vertices
\begin{equation}\label{cubicv}
    {\cal V}_{12 \phi} = g_{12}\, \phi_1 \phi_2 \varphi, \qquad {\cal V}_{34 \phi} = g_{34}\, \phi_3 \phi_4 \varphi.
\end{equation}
In the following we shall take for simplicity the external scalar fields $\phi_i$ to be massless, with the case of external massive fields discussed in appendix \ref{A::MassiveExt}.\footnote{Like for the contact diagrams in the previous section, each external massive field introduces a Mellin-Barnes integral \eqref{SPm} encoding the dependence on the particle mass.} The exchange diagram with external massless scalars reads 
\begin{multline}\label{exchdef}
   {\cal A}^{{\cal V}_{12 \varphi}{\cal V}_{34 \varphi}}\left(Y_1,Y_2,Y_3,Y_4\right) = (-i g_{12})(-i g_{34}) \int {\rm d}^{d+2}X {\rm d}^{d+2}Y\, G^{(0)}_{T}\left(X,Y_1\right)G^{(0)}_{T}\left(X,Y_2\right) \\ \times G^{(m)}_{T}\left(X,Y\right)G^{(0)}_{T}\left(Y,Y_3\right)G^{(0)}_{T}\left(Y,Y_4\right).
\end{multline}
Employing Schwinger parametrization, which for the massless scalars is simply \eqref{MLschw}, this is given by 
\begin{multline}\label{GaussExc}
    {\cal A}^{{\cal V}_{12 \varphi}{\cal V}_{34 \varphi}}\left(Y_1,Y_2,Y_3,Y_4\right)
    = - g_{12}g_{34} \left(\frac{1}{4\pi^{\frac{d+2}{2}}}i^{-\frac{d}{2}}\right)^4\, \left(\frac{1}{4\pi^{\frac{d+2}{2}}}\left(\frac{m}{2}\right)^{\frac{d}{2}}\right) \\ \times  \int^{+i\infty}_{-i\infty}\frac{{\rm d}{\bar s}}{2\pi i}\Gamma\left({\bar s}-\tfrac{d}{4}\right)\left(\frac{m}{2}\right)^{-2{\bar s}} i^{-({\bar s}+\frac{d}{4})} \\ \times \int^{\infty}_0 \frac{{\rm d}{\bar t}}{{\bar t}} {\bar t}^{{\bar s}+\frac{d}{4}} \prod^4_{i=1} \frac{{\rm d}t_i}{t_i} t^{\frac{d}{2}}_i\, \int {\rm d}^{d+2}X {\rm d}^{d+2}Y\, \exp\left[i {\bar t}\left(X-Y\right)^2+i t_1\left(X-Y_1\right)^2+i t_2\left(X-Y_2\right)^2 \right. \\ \left. +i t_3\left(Y-Y_3\right)^2+i t_4\left(Y-Y_4\right)^2\right],
\end{multline}
where the integral over the internal points is reduced to two nested Gaussian integrals (see appendix \ref{A::GI}).

\vskip 4pt
The contribution to the corresponding celestial correlation function is then:
\begin{align}\label{Cexch}
   {\cal A}^{{\cal V}_{12 \varphi}{\cal V}_{34 \varphi}}_{\Delta_1 \Delta_2\Delta_3\Delta_4}\left(Q_1,Q_2,Q_3,Q_4\right)&= \lim_{{\hat Y}_i \to Q_i}   \int^{\infty}_0 \left(\prod\limits^4_{i=1} \frac{{\rm d}R_i}{R_i} R^{\Delta_i}_i\right) {\cal A}^{{\cal V}_{12 \varphi}{\cal V}_{34 \varphi}}\left(Y_1,Y_2,Y_3,Y_4\right)\\\nonumber &=- g_{12}g_{34}i^{d}\pi^{d+2}\left(\frac{1}{4} \frac{1}{\pi^{\frac{d+2}{2}}}i^{-\frac{d}{2}}\right)^4\, \left(\frac{1}{4} \frac{1}{\pi^{\frac{d+2}{2}}}\left(\frac{m}{2}\right)^{\frac{d}{2}}\right)\\ \nonumber & \times  \int^{\infty}_0 \left( \prod\limits^4_{i=1} \frac{{\rm d}R_i}{R_i} R^{\Delta_i}_i \right)\int^{+i\infty}_{-i\infty}\frac{{\rm d}{\bar s}}{2\pi i}\Gamma\left({\bar s}-\tfrac{d}{4}\right)\left(\frac{m}{2}\right)^{-2{\bar s}} i^{-({\bar s}+\frac{d}{4})}\\ \nonumber & \times \int^{\infty}_0 \frac{{\rm d}{\bar t}}{{\bar t}} {\bar t}^{{\bar s}+\frac{d}{4}} \prod^4_{i=1} \frac{{\rm d}t_i}{t_i} t^{\frac{d}{2}}_i\,\left[({\bar t}+t_1+t_2)({\bar t}+t_3+t_4)-{\bar t}^2\right]^{-\frac{d+2}{2}}\\ \nonumber & \hspace*{-4cm}\times \exp \left[-2i\frac{({\bar t}+t_3+t_4)t_1t_2R_1 R_2 Q_1 \cdot Q_2+({\bar t}+t_1+t_2)t_3t_4R_3 R_4 Q_3 \cdot Q_4}{({\bar t}+t_1+t_2)({\bar t}+t_3+t_4)-{\bar t}^2}\right. \\ \nonumber & \hspace*{-4cm} \left.-2i \frac{{\bar t}\,t_1t_3 R_1 R_3 Q_1 \cdot Q_3+{\bar t}\,t_2t_3 R_2 R_3 Q_2 \cdot Q_3+{\bar t}\,t_1t_4 R_1 R_4 Q_1 \cdot Q_4+{\bar t}\,t_2t_4 R_2 R_4 Q_2 \cdot Q_4}{({\bar t}+t_1+t_2)({\bar t}+t_3+t_4)-{\bar t}^2}\right].
\end{align}
The Mellin-Barnes representation \eqref{expMB} of the exponential then gives the Mellin amplitude:  
\begin{multline}\label{mellinAmplexch}
    M^{{\cal V}_{12 \varphi}{\cal V}_{34 \varphi}}_{\Delta_1 \Delta_2\Delta_3\Delta_4}\left(\delta_{ij}\right)=- g_{12}g_{34}\,i^{d}i^{+\frac{1}{2}\sum_i \Delta_i}\pi^{d+2}\left(\frac{1}{4} \frac{1}{\pi^{\frac{d+2}{2}}}i^{-\frac{d}{2}}\right)^4\, \left(\frac{1}{4} \frac{1}{\pi^{\frac{d+2}{2}}}\left(\frac{m}{2}\right)^{\frac{d}{2}}\right)\\ \times \int^{+i\infty}_{-i\infty}\frac{{\rm d}{\bar s}}{2\pi i}\Gamma\left({\bar s}-\tfrac{d}{4}\right)\left(\frac{m}{2}\right)^{-2{\bar s}} i^{-({\bar s}+\frac{d}{4})} \\
    T_{\text{exch}}\left(\tfrac{d}{2}-\Delta_1,\tfrac{d}{2}-\Delta_2,\tfrac{d}{2}-\Delta_3,\tfrac{d}{2}-\Delta_4;{\bar s}+\tfrac{d}{4};\tfrac{\beta}{2};0,0\right),
\end{multline}
where for exchanges the integrals over the Schwinger parameters take the form:\footnote{In the case of external massive scalar fields the integrals over the Schwinger parameters take the same form -- see appendix \ref{A::MassiveExt}.}
\begin{multline}
T_{\text{exch}}\left(a_1,a_2,a_3,a_4;a;b;b_{12},b_{34}\right)=\int^\infty_0 \frac{{\rm d}{\bar t}}{{\bar t}} {\bar t}^{-(\delta_{13}+\delta_{14}+\delta_{23}+\delta_{24})+a}\\ \times \int^\infty_0 \prod^4_{i=1}\frac{{\rm d}t_i}{t_i} t^{a_i}_i\left[({\bar t}+t_1+t_2)({\bar t}+t_3+t_4)-{\bar t}^2\right]^{b}\\ \times \left({\bar t}+t_1+t_2\right)^{-\delta_{34}+b_{12}}\left({\bar t}+t_3+t_4\right)^{-\delta_{12}+b_{34}}.
\end{multline}
Evaluating the integrals following the steps in appendix \ref{A::SI} gives the result in terms of the generalised hypergeometric function:
\begin{multline}\label{mastintegr}
    T_{\text{exch}}\left(a_1,a_2,a_3,a_4;a;b;b_{12},b_{34}\right) = 2\pi i\, \delta \left(a+a_1+a_2+a_3+a_4+2b+b_{12}+b_{34}-\tfrac{d+2}{2}-\tfrac{\beta}{2}\right)\\ \times \left(\prod^4_{i=1}\Gamma\left(a_i\right)\right) \frac{\Gamma\left(\delta_{34}-b-b_{12}-a_1-a_2\right)\Gamma\left(\delta_{12}-b-b_{34}-a_3-a_4\right)}{\Gamma\left(\delta_{34}-b-b_{12}\right)\Gamma\left(\delta_{12}-b-b_{34}\right)}\\ \times  {}_3F_2\left(\begin{matrix}\delta_{12}-b-b_{34}-a_3-a_4,\delta_{34}-b-b_{12}-a_1-a_2,-b\\\delta_{12}-b-b_{34},\delta_{34}-b-b_{12} \end{matrix};1\right),
\end{multline}
where the Dirac delta function allows to eliminate the integral over ${\bar s}$ associated to the mass of the exchanged field. This gives the Mellin amplitude: 
\begin{shaded}
  \begin{multline}\label{MAexchdij}
    M^{{\cal V}_{12 \varphi}{\cal V}_{34 \varphi}}_{\Delta_1 \Delta_2\Delta_3\Delta_4}\left(\delta_{ij}\right)=g_{12}g_{34}\,\pi^{d+2}\,\prod^4_{i=1}\frac{1}{4\pi^{\frac{d+2}{2}}}\Gamma\left(\tfrac{d}{2}-\Delta_i\right)\\ \times \frac{1}{4\pi^{\frac{d+2}{2}}} \left(\frac{m}{2}\right)^{3d-4-\sum_i\Delta_i}\Gamma\left(\frac{-3d+4+\sum_i\Delta_i}{2}\right)\\ \times \frac{\Gamma\left(\delta_{12}-\frac{\beta}{2}+\Delta_3+\Delta_4-d\right)\Gamma\left(\delta_{34}-\frac{\beta}{2}+\Delta_1+\Delta_2-d\right)}{\Gamma\left(\delta_{12}-\frac{\beta}{2}\right)\Gamma\left(\delta_{34}-\frac{\beta}{2}\right)}\\
    {}_3F_2\left(\begin{matrix}\delta_{12}-\frac{\beta}{2}+\Delta_3+\Delta_4-d,\delta_{34}-\frac{\beta}{2}+\Delta_1+\Delta_2-d,-\frac{\beta}{2}\\\delta_{12}-\frac{\beta}{2},\delta_{34}-\frac{\beta}{2} \end{matrix};1\right).
\end{multline}  
\end{shaded}

\begin{figure}[t]
    \centering
    \includegraphics[width=0.8\textwidth]{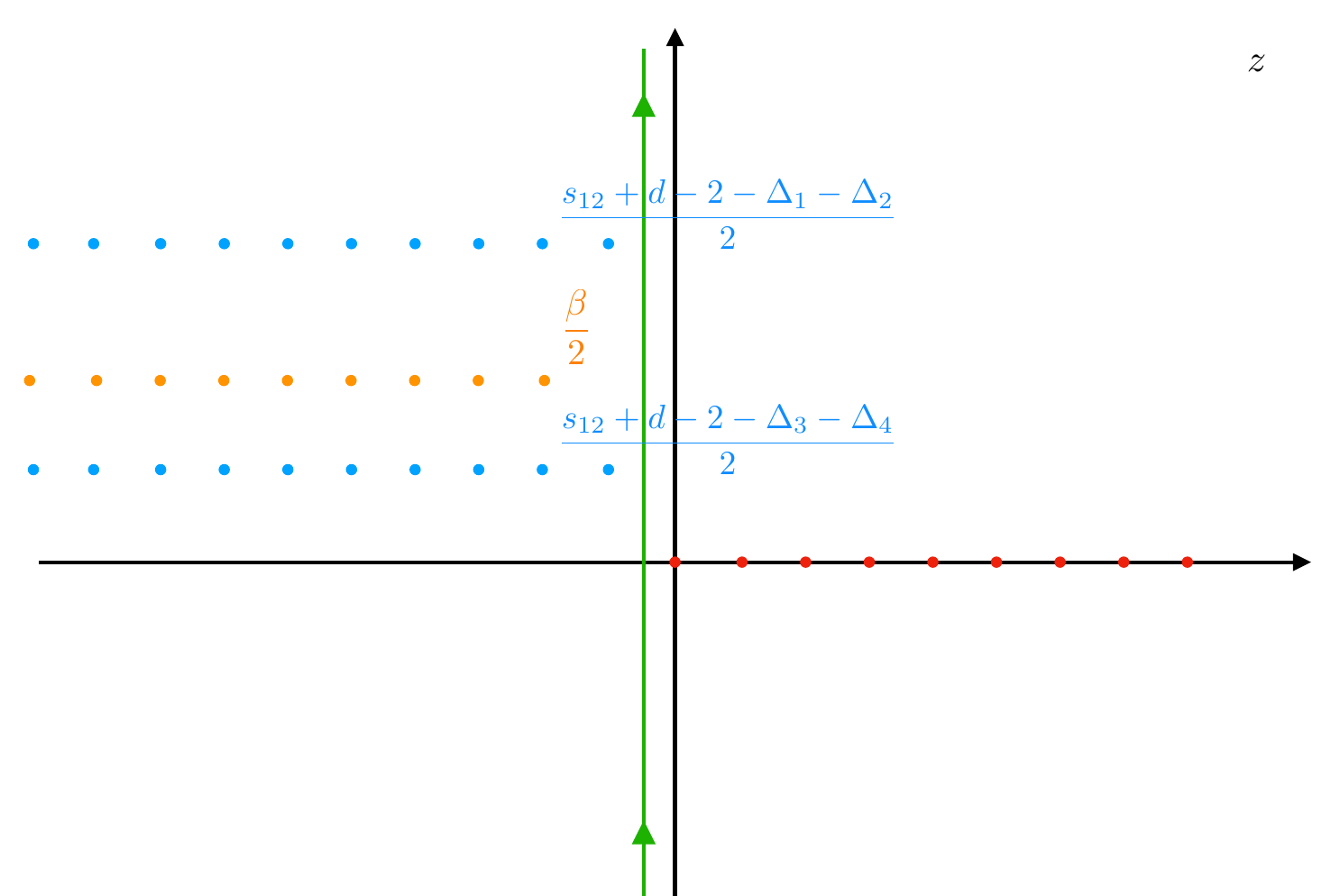}
    \caption{Poles of the Mellin-Barnes integral \eqref{Mexchs12s13} in the complex $z$ plane. The poles \eqref{ppexch} of the Mellin amplitude in $s_{12}$ are generated when the integration contour (green) gets pinched between a blue pole and a red pole.}
    \label{fig::PPexch}
\end{figure}

\begin{figure}[t]
    \centering
    \includegraphics[width=0.8\textwidth]{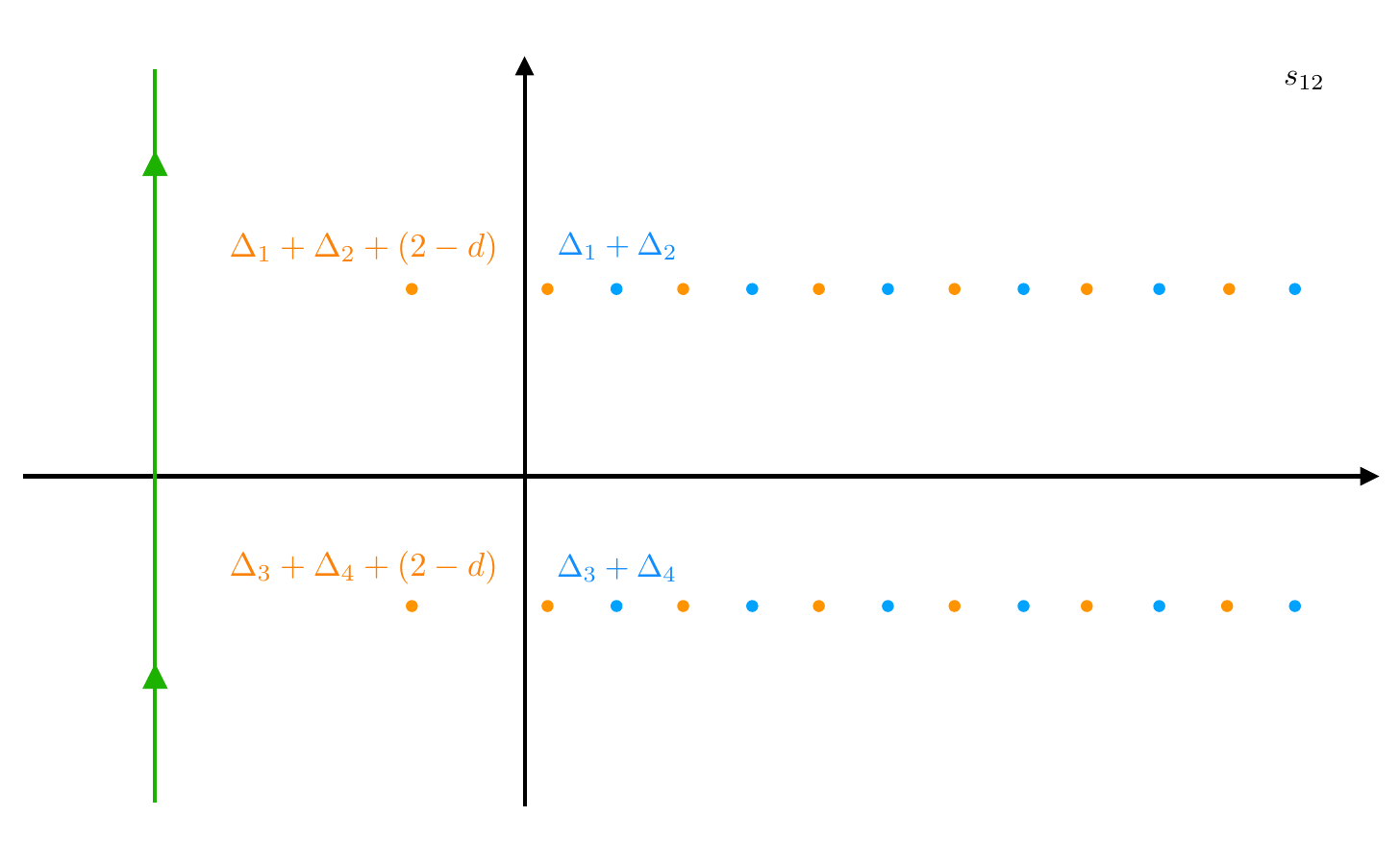}
    \caption{Poles in the the Mellin representation of the Celestial exchange diagram \eqref{exchdef}. Yellow poles are the poles of the corresponding Mellin amplitude \eqref{Mexchs12s13}, encoding the particle exchange. Blue poles are the ``double-trace poles" \eqref{DTpoles0} in the integrand of the Mellin representation \eqref{MBst}.}
    \label{fig::exch_poles}
\end{figure}

To study the pole structure of the Mellin amplitude, it is useful to employ the Mellin-Barnes representation of the generalised hypergeometric function:
\begin{multline}
    {}_3F_2\left(\begin{matrix}a,b,c\\e,f \end{matrix};1\right) = \frac{\Gamma\left(e\right)\Gamma\left(f\right)}{\Gamma\left(a\right)\Gamma\left(b\right)\Gamma\left(c\right)} \int^{+i\infty}_{-i\infty} \frac{{\rm d}z}{2\pi i}\frac{\Gamma\left(z+a\right)\Gamma\left(z+b\right)\Gamma\left(z+c\right)}{\Gamma\left(z+e\right)\Gamma\left(z+f\right)}\\ \times \Gamma\left(-z\right)\left(-1\right)^z.
\end{multline}
Expressing the Mellin amplitude in terms of the ``Mandelstam invariants" \eqref{MBmandel}, this brings the four-point function into the form \eqref{MBst} with: 
\begin{multline}\label{Mexchs12s13}
    M^{{\cal V}_{12 \varphi}{\cal V}_{34 \varphi}}_{\Delta_1 \Delta_2\Delta_3\Delta_4}\left(s_{12},s_{13}\right)=g_{12}g_{34}\,\pi^{d+2}\,\prod^4_{i=1}\frac{1}{4\pi^{\frac{d+2}{2}}}\Gamma\left(\tfrac{d}{2}-\Delta_i\right)\,\\ \times \frac{1}{4\pi^{\frac{d+2}{2}}} \left(\frac{m}{2}\right)^{3d-4-\sum_i\Delta_i}\Gamma\left(\frac{-3d+4+\sum_i\Delta_i}{2}\right)\\ \times \frac{1}{\Gamma\left(-\frac{\beta}{2}\right)}\int^{+i\infty}_{-i\infty}\frac{{\rm d}z}{2\pi i}\frac{\Gamma\left(z+\tfrac{1}{2}(2-d+\Delta_1+\Delta_2-s_{12})\right)\Gamma\left(z+\tfrac{1}{2}(2-d+\Delta_3+\Delta_4-s_{12})\right)}{\Gamma\left(z+\tfrac{1}{2}(d+2-\Delta_3-\Delta_4-s_{12})\right)\Gamma\left(z+\tfrac{1}{2}(d+2-\Delta_1-\Delta_2-s_{12})\right)}\\ \times \Gamma\left(z-\tfrac{\beta}{2}\right)\Gamma\left(-z\right)\left(-1\right)^z.
\end{multline}
Note that this only depends on the variable $s_{12}$, as expected for Mellin amplitudes in scalar field theories. The poles in $s_{12}$ can be extracted from the above expression by studying the pole pinching of the $z$ integral (see figure \ref{fig::PPexch}), which leads to the following double-trace-like singularities (where $n=0,1,2, \ldots\,$):
\begin{equation}\label{ppexch}
    s_{12} = \Delta_1+\Delta_2+(2-d)+2n, \qquad s_{12} = \Delta_3+\Delta_4+(2-d)+2n.
\end{equation}
These are in addition to the families of genuine double-trace poles arising from the Mellin transform \eqref{MBst} of conformal correlators, which we repeat below for convenience (where $n=0,1,2, \ldots\,$):
\begin{align}\label{DTexch}
    s_{12} = \Delta_1+\Delta_2+2n, \qquad s_{12} = \Delta_3+\Delta_4+2n.
\end{align}
These correspond to contributions from double-trace operators \eqref{DTops} with spin-$\ell=0$ and their descendants. Notice that for even dimensions $d$ the double-trace-like poles \eqref{ppexch} mix with the double-trace poles \eqref{DTexch}, which generates anomalous dimensions of the double-trace operators \eqref{DTops} already at tree level, which in AdS-CFT only occurs when $\Delta_1+\Delta_2=\Delta_3+\Delta_4$. The pole structure of the exchange Mellin amplitude is summarised in figure \ref{fig::exch_poles}. 

\vskip 4pt
All other singularities of the $z$ integral \eqref{Mexchs12s13} are independent of the Mellin variables $s_{12}$ and $s_{13}$. From pole pinching of the integration contour (see figure \ref{fig::PPexch}), there are also simple poles at
\begin{equation}
    \beta = 2n, \quad n=0, 1, 2, \ldots\,.
\end{equation}
These are canceled by the factor $1/\Gamma\left(-\frac{\beta}{2}\right)$ in the expression \eqref{Mexchs12s13} for the Mellin amplitude. Another family of poles can be seen by analysing the convergence of the $z$ integral. In particular, setting $z=Re^{i\theta}$ the behaviour of the integrand as $R \to \infty$ is
\begin{equation}
    \sim e^{-\pi  R (| \sin (\theta )| +\sin (\theta ))}\,R^{\text{Re}\left[\frac{1}{2}\left(-3d+\sum_i\Delta_i\right)\right]}.
\end{equation}
There is therefore a divergence for $\text{Re}\left[\frac{1}{2}\left(-3d+\sum_i\Delta_i\right)\right] \geq -1$ and a corresponding family of poles at
\begin{equation}
    \frac{-3d+4+\sum_i\Delta_i}{2}=1+n, \quad n=0, 1, 2, \ldots\,.
\end{equation}

\vskip 4pt
In the above discussion we focused on correlators with external massless scalar fields. Exchange Mellin amplitudes involving massive external fields and their poles in $s_{12}$ are discussed in appendix \ref{A::MassiveExt}.

\vskip 4pt
\paragraph{Massless exchanged scalar.} Let us now consider the case that the exchanged scalar is also massless. This can be obtained from the limit $m\to 0$ of the massive exchange above, since the massless limit of the Feynman propagator is smooth. To this end one uses that:
\begin{equation}
\lim_{m\to0} \left(\frac{m}{2}\right)^{3d-4-\sum_i\Delta_i}\Gamma\left(\frac{-3d+4+\sum_i\Delta_i}{2}\right) = 2 \pi i\, \delta\left(\frac{-3d+4+\sum_i\Delta_i}{2}\right),
\end{equation}
which gives the Mellin amplitude:
\begin{multline}\label{Mexchs12s13ml}
    M^{{\cal V}_{12 \varphi}{\cal V}_{34 \varphi}}_{\Delta_1 \Delta_2\Delta_3\Delta_4}\left(s_{12},s_{13}\right)=g_{12}g_{34}\,\pi^{d+2}\,\prod^4_{i=1}\frac{1}{4\pi^{\frac{d+2}{2}}}\Gamma\left(\tfrac{d}{2}-\Delta_i\right)\,\\ \times \frac{1}{4\pi^{\frac{d+2}{2}}}2\pi i\, \delta\left(\frac{-3d+4+\sum_i\Delta_i}{2}\right) \frac{\Gamma\left(\tfrac{2-d+\Delta_1+\Delta_2-s_{12}}{2}\right)\Gamma\left(\tfrac{2-d+\Delta_3+\Delta_4-s_{12}}{2}\right)}{\Gamma\left(\tfrac{d+2-\Delta_1-\Delta_2-s_{12}}{2}\right)\Gamma\left(\tfrac{d+2-\Delta_3-\Delta_4-s_{12}}{2}\right)}\\ \times {}_3F_2\left(\begin{matrix}\tfrac{2-d+\Delta_1+\Delta_2-s_{12}}{2},\tfrac{2-d+\Delta_3+\Delta_4-s_{12}}{2},-\frac{\beta}{2}\\\tfrac{d+2-\Delta_1-\Delta_2-s_{12}}{2},\tfrac{d+2-\Delta_3-\Delta_4-s_{12}}{2} \end{matrix};1\right).
\end{multline}
This is equal to the expression that one would obtain by using the massless Feynman propagator \eqref{Spml} for the exchanged field from the beginning (as expected).

\vskip 4pt
Like for the contact diagram \eqref{MAcml} for massless scalars, there is a delta-function constraint on the scaling dimensions when the exchanged field is massless as well. This leads to some simplifications with respect to the massive case \eqref{Mexchs12s13}. In particular, on the support of the Dirac delta function we have $\beta = 2\left(d-3\right)$ and for integer dimensions $d \geq 3$ we can use Saalsch\"utz's theorem:
\begin{equation}\label{ssth}
    {}_3F_2\left(\begin{matrix}a,b,-n\\c,1+a+b-c-n\end{matrix};1\right) = \frac{\left(c-a\right)_n\left(c-b\right)_n}{\left(c\right)_n\left(c-a-b\right)_n}.
\end{equation}
This gives the following expression for the Mellin amplitude (for integer dimensions $d \geq 3$):
\begin{multline}\label{mldge3}
    M^{{\cal V}_{12 \varphi}{\cal V}_{34 \varphi}}_{\Delta_1 \Delta_2\Delta_3\Delta_4}\left(s_{12},s_{13}\right)=g_{12}g_{34}\,\pi^{d+2}\,\prod^4_{i=1}\frac{1}{4\pi^{\frac{d+2}{2}}}\Gamma\left(\tfrac{d}{2}-\Delta_i\right)\,\\ \times (-1)^{3-d} \frac{\Gamma\left(\frac{d}{2}-1\right)}{\Gamma\left(2-\frac{d}{2}\right)} \frac{\Gamma\left(2d-3-\Delta_1-\Delta_2\right)}{\Gamma\left(d-\Delta_1-\Delta_2\right)} \\ \times \frac{1}{4\pi^{\frac{d+2}{2}}}2\pi i\, \delta\left(\frac{-3d+4+\sum_i\Delta_i}{2}\right) \frac{\Gamma \left(\frac{2-d+\Delta_3+\Delta_4-s_{12}}{2}\right) \Gamma \left(\frac{2-d+\Delta_1+\Delta_2-s_{12}}{2}\right)}{\Gamma\left(\frac{\Delta_3+\Delta_4-s_{12}}{2}\right)\Gamma\left(\frac{\Delta_1+\Delta_2-s_{12}}{2}\right)}.
\end{multline}

Let us make some observations:

\begin{itemize}
    \item Note that the double-trace poles \eqref{DTexch} in the Mellin integrand are completely canceled by those in the denominator of the Mellin amplitude \eqref{mldge3}. The only contributions for $d \geq 3$ are therefore from the poles \eqref{ppexch}:
\begin{align}
   s_{12} = \Delta_1+\Delta_2+(2-d)+2n, \quad s_{12} = \Delta_3+\Delta_4+(2-d)+2n.
\end{align}

\item The Mellin amplitude \eqref{mldge3} is proportional to that of a spin-$0$ conformal partial wave (see e.g. \cite{Sleight:2018epi} equation (3.6)): 
\begin{equation}\label{cpw}
        {\cal F}_{\Delta,0}\left(s_{12},s_{13}\right) \propto \frac{\Gamma\left(\frac{\Delta-s_{12}}{2}\right)\Gamma\left(\frac{d-\Delta-s_{12}}{2}\right)}{\Gamma\left(\frac{\Delta_1+\Delta_2-s_{12}}{2}\right)\Gamma\left(\frac{\Delta_3+\Delta_4-s_{12}}{2}\right)}.
    \end{equation} 
A conformal partial wave ${\cal F}_{\Delta,0}$ is a (single-valued) sum of two conformal blocks which encode the contributions from the conformal multiplet with a primary operator of scaling dimension $\Delta$ and the shadow multiplet with scaling dimension $d-\Delta$. In the exchange \eqref{mldge3}, by virtue of the Dirac delta function constraint, we have:
\begin{equation}\label{dtrad}
\Delta=\Delta_1+\Delta_2+(2-d), \qquad d-\Delta=\Delta_3+\Delta_4+(2-d).
\end{equation}
The exchange  \eqref{mldge3} is therefore a sum of two conformal blocks, one for each of the scaling dimensions \eqref{dtrad}. This feature moreover extends to exchanges of massless fields with non-trivial spin, which will be presented in the upcoming work \cite{toappear}.

\item The Mellin amplitude \eqref{mldge3} is vanishing for even dimensions $d>2$, which is consistent with the vanishing \eqref{mlnptvanish} of the corresponding three-point contact diagrams of massless scalars in this case for non-null separated points $Q_i \cdot Q_j \ne 0$.
\end{itemize}

\vskip 4pt
For $d=2$ we have instead:
\begin{multline}\label{mlexchd2}
    M^{{\cal V}_{12 \varphi}{\cal V}_{34 \varphi}}_{\Delta_1 \Delta_2\Delta_3\Delta_4}\left(s_{12},s_{13}\right)=g_{12}g_{34}\,\pi^{d+2}\,\prod^4_{i=1}\frac{1}{4\pi^{\frac{d+2}{2}}}\Gamma\left(\tfrac{d}{2}-\Delta_i\right)\, \\ \times \frac{1}{4\pi^{\frac{d+2}{2}}}2\pi i\, \delta\left(\frac{-3d+4+\sum_i\Delta_i}{2}\right) \\ \times \frac{\psi\left(\frac{\Delta_1+\Delta_2-s_{12}}{2} \right)-\psi\left(\frac{\Delta_3+\Delta_4-s_{12}}{2}\right)}{\Delta_1+\Delta_2-1},
\end{multline}
where $\psi(z)$ is the polygamma function $\psi\left(z\right)=\Gamma^\prime(z)/\Gamma(z)$, which has simple poles for non-positive integers $z$. For $d=2$ the double-trace \eqref{DTexch} and double-trace-like poles \eqref{ppexch} overlap completely, giving only double poles at the values:
\begin{align}
    s_{12} = \Delta_1+\Delta_2+2n, \qquad s_{12} = \Delta_3+\Delta_4+2n.
\end{align}

\subsection{Loops}
\label{subsec::1loop}

\begin{figure}[t]
    \centering
    \includegraphics[width=0.3\textwidth]{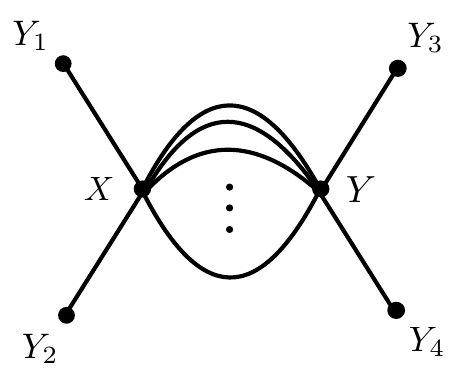}
    \caption{Four-point loop diagram generated by the vertices \eqref{nptv}.}
    \label{fig::higherloop}
\end{figure}

The approach outlined at the beginning of section \ref{sec::MA4pcc} extends straightforwardly to loop diagrams. In the following this is illustrated for four-point loop diagrams with two internal points generated by vertices (see figure \ref{fig::higherloop})
\begin{equation}\label{nptv}
    {\cal V}_{12 \varphi_1 \ldots \varphi_n} = g_{12}\, \phi_1 \phi_2 \varphi_1 \ldots \varphi_n, \qquad {\cal V}_{34 \varphi_1 \ldots \varphi_n} = g_{34}\, \phi_3 \phi_4 \varphi_1 \ldots \varphi_n,
\end{equation}
with $n>1$. The case $n=1$ corresponds to the tree-level exchange diagram \eqref{exchdef}. For ease of illustration we will take the external fields $\phi_i$ to be massless and internal fields $\varphi_i$ with mass $m_i$, though the extension to massive external fields is straightforward (see appendix \ref{A::MassiveExt}).

\paragraph{One-loop.} The simplest case is $n=2$. This is given by
\begin{multline}
   {\cal A}^{{\cal V}_{12 \varphi_1 \varphi_2}{\cal V}_{34 \varphi_1 \varphi_2}}\left(Y_1,Y_2,Y_3,Y_4\right) = (-i g_{12})(-i g_{34}) \int {\rm d}^{d+2}X {\rm d}^{d+2}Y\, G^{(0)}_{T}\left(X,Y_1\right)G^{(0)}_{T}\left(X,Y_2\right) \\ \times G^{(m_1)}_{T}\left(X,Y\right)G^{(m_2)}_{T}\left(X,Y\right)G^{(0)}_{T}\left(Y,Y_3\right)G^{(0)}_{T}\left(Y,Y_4\right).
\end{multline}
Employing Schwinger parametrization we have,
\begin{multline}
  \hspace*{-0.6cm}  {\cal A}^{{\cal V}_{12 \varphi_1 \varphi_2}{\cal V}_{34 \varphi_1 \varphi_2}}\left(Y_1,Y_2,Y_3,Y_4\right)
    = - g_{12}g_{34} \left(\frac{1}{4} \frac{1}{\pi^{\frac{d+2}{2}}}i^{-\frac{d}{2}}\right)^4\, \left(\frac{1}{4} \frac{1}{\pi^{\frac{d+2}{2}}}\left(\frac{m_1}{2}\right)^{\frac{d}{2}}\right)\left(\frac{1}{4} \frac{1}{\pi^{\frac{d+2}{2}}}\left(\frac{m_2}{2}\right)^{\frac{d}{2}}\right) \\ \times  \int^{+i\infty}_{-i\infty}\frac{{\rm d}{\bar s}_1}{2\pi i}\frac{{\rm d}{\bar s}_2}{2\pi i}\Gamma\left({\bar s}_1-\tfrac{d}{4}\right)\Gamma\left({\bar s}_2-\tfrac{d}{4}\right)\left(\frac{m_1}{2}\right)^{-2{\bar s}_1}\left(\frac{m_2}{2}\right)^{-2{\bar s}_2} i^{-({\bar s}_1+\frac{d}{4})}i^{-({\bar s}_2+\frac{d}{4})} \\ \times \int^{\infty}_0 \frac{{\rm d}{\bar t}_1}{{\bar t}_1}\frac{{\rm d}{\bar t}_2}{{\bar t}_2} {\bar t}_1^{{\bar s}_1+\frac{d}{4}} {\bar t}_2^{{\bar s}_2+\frac{d}{4}} \prod^4_{i=1} \frac{dt_i}{t_i} t^{\frac{d}{2}}_i\, \int {\rm d}^{d+2}X {\rm d}^{d+2}Y\, \exp\left[i t_1\left(X-Y_1\right)^2+i t_2\left(X-Y_2\right)^2 \right. \\ \left.+ i ({\bar t}_1+{\bar t}_2)\left(X-Y\right)^2+i t_3\left(Y-Y_3\right)^2+i t_4\left(Y-Y_4\right)^2\right],
\end{multline}
where the Gaussian integral is the same as for the exchange diagram \eqref{exchdef}, but with the Schwinger parameter associated to the exchanged field replaced by the sum of the Schwinger parameters ${\bar t}_1$ and ${\bar t}_2$ for the two exchanged particles. 

\vskip 4pt
It then immediately follows that the Mellin amplitude for the corresponding contribution to the celestial four-point function is given by:
\begin{multline}
 \hspace*{-0.75cm}   M^{{\cal V}_{12 \varphi_1 \varphi_2}{\cal V}_{34 \varphi_1 \varphi_2}}_{\Delta_1 \Delta_2\Delta_3\Delta_4}\left(\delta_{ij}\right)=- g_{12}g_{34} i^{-d} \pi^{d+2}\, \left( \frac{1}{4\pi^{\frac{d+2}{2}}}\left(\frac{m_1}{2}\right)^{\frac{d}{2}}\right)\left( \frac{1}{4\pi^{\frac{d+2}{2}}}\left(\frac{m_2}{2}\right)^{\frac{d}{2}}\right)\prod^4_{i=1}\frac{i^{\frac{\Delta_i}{2}}}{4\pi^{\frac{d+2}{2}}}\Gamma\left(\tfrac{d}{2}-\Delta_i\right) \\ \times  \int^{+i\infty}_{-i\infty}\frac{{\rm d}{\bar s}_1}{2\pi i}\frac{{\rm d}{\bar s}_2}{2\pi i}\Gamma\left({\bar s}_1-\tfrac{d}{4}\right)\Gamma\left({\bar s}_2-\tfrac{d}{4}\right)\left(\frac{m_1}{2}\right)^{-2{\bar s}_1}\left(\frac{m_2}{2}\right)^{-2{\bar s}_2} i^{-({\bar s}_1+\frac{d}{4})}i^{-({\bar s}_2+\frac{d}{4})} \\ \times \int^{\infty}_0 \frac{{\rm d}{\bar t}_1}{{\bar t}_1}\frac{{\rm d}{\bar t}_2}{{\bar t}_2} {\bar t}_1^{{\bar s}_1+\frac{d}{4}} {\bar t}_2^{{\bar s}_2+\frac{d}{4}} \left({\bar t}_1+{\bar t}_2\right)^{d-2-\tfrac{1}{2}\sum_i\Delta_i} \\ \times \frac{1}{\Gamma\left(-\tfrac{\beta}{2}\right)}\int^{+i\infty}_{-i\infty}\frac{{\rm d}z}{2\pi i}\frac{\Gamma\left(\delta_{34}+z-\tfrac{\beta}{2}+\Delta_1+\Delta_2-d\right)\Gamma\left(\delta_{12}+z-\tfrac{\beta}{2}+\Delta_3+\Delta_4-d\right)}{\Gamma\left(\delta_{12}+z-\tfrac{\beta}{2}\right)\Gamma\left(\delta_{34}+z-\tfrac{\beta}{2}\right)}\\ \times \Gamma\left(z-\tfrac{\beta}{2}\right)\Gamma\left(-z\right)\left(-1\right)^z,
\end{multline}
where the integrals over the Schwinger parameters $t_i$ associated to the external legs were the same as those encountered for the four-point exchange, with the parameter ${\bar t}$ associated to the exchanged particle is replaced by the sum of parameters ${\bar t}_1+{\bar t}_2$ for the two exchanged fields. The integrals over the remaining Schwinger parameters ${\bar t}_1$ and ${\bar t}_2$ take the same form as those encountered in the Mellin amplitude \eqref{mellinAmplCont} for contact diagrams:
\begin{multline}
    \int^{\infty}_0 \frac{{\rm d}{\bar t}_1}{{\bar t}_1}\frac{{\rm d}{\bar t}_2}{{\bar t}_2} {\bar t}_1^{{\bar s}_1+\frac{d}{4}} {\bar t}_2^{{\bar s}_2+\frac{d}{4}} \left({\bar t}_1+{\bar t}_2\right)^{d-2-\tfrac{1}{2}\sum_i\Delta_i} \\= \frac{\Gamma\left({\bar s}_1+\frac{d}{4}\right)\Gamma\left({\bar s}_2+\frac{d}{4}\right)}{\Gamma\left(\tfrac{4-2d+\sum_i\Delta_i}{2}\right)} \left(2\pi i\right) \delta \left(\tfrac{-3d+4-2({\bar s}_1+{\bar s}_2)+\sum_i\Delta_i}{2}\right).
\end{multline}
What remains are the Mellin-Barnes integrals in the variables ${\bar s}_1$ and ${\bar s}_2$, which encode the dependence on the masses $m_1$ and $m_2$ of the two particles running in the loop. These are given by Gauss hypergeometric functions:\footnote{The second equality follows from the transformation formula:
\begin{multline}{}_2F_1\left(a,b;c;z\right)=\frac{\Gamma(c)\Gamma(a+b-c)}{\Gamma(a)\Gamma(b)}{}_2F_1\left(c-a,c-b;c-a-b+1;1-z\right)(1-z)^{c-a-b}\\+\frac{\Gamma(c)\Gamma(c-a-b)}{\Gamma(c-a)\Gamma(c-b)}{}_2F_1\left(a,b;a+b-c+1;1-z\right).\end{multline}}
\begin{multline}
    \frac{1}{\Gamma\left(\tfrac{4-2d+\sum_i\Delta_i}{2}\right) \Gamma \left(\tfrac{4-3d+\sum_i \Delta_i}{2}\right)} \int^{+i\infty}_{-i\infty}\frac{{\rm d}{\bar s}_1}{2\pi i}\frac{{\rm d}{\bar s}_2}{2\pi i}\left(2\pi i\right) \delta \left(\tfrac{-3d+4-2({\bar s}_1+{\bar s}_2)+\sum_i\Delta_i}{2}\right) \\ \times \Gamma\left({\bar s}_1+\tfrac{d}{4}\right)\Gamma\left({\bar s}_2+\tfrac{d}{4}\right)\Gamma\left({\bar s}_1-\tfrac{d}{4}\right)\Gamma\left({\bar s}_2-\tfrac{d}{4}\right)\left(\frac{m_1}{2}\right)^{-2{\bar s}_1}\left(\frac{m_2}{2}\right)^{-2{\bar s}_2} \\
= \left(\frac{m_1}{2}\right)^{-4+3d-\sum_i\Delta_i} \left[ \left(\frac{m_2}{m_1}\right)^{\frac{d}{2}} \Gamma \left(-\frac{d}{2}\right) {}_2F_1\left(\frac{4-3 d+\sum_i\Delta_i}{2},\frac{4-2d+\sum_i\Delta_i}{2};\frac{d+2}{2};\frac{m^2_2}{m_1^2}\right) \right. \\ \left.
+\left(\frac{m_2}{m_1}\right)^{-\frac{d}{2}} \frac{\Gamma\left(\tfrac{4-4d+\sum_i\Delta_i}{2}\right)}{\Gamma\left(\tfrac{4-2d+\sum_i\Delta_i}{2}\right)} \Gamma \left(\frac{d}{2}\right)  {}_2F_1\left(\frac{4-3d+\sum_i\Delta_i}{2},\frac{4-4d+\sum_i\Delta_i}{2};\frac{2-d}{2};\frac{m^2_2}{m_1^2}\right)\right]\\
=\left(\frac{m_1}{2}\right)^{-4+3d-\sum_i\Delta_i} \left(\frac{m_2}{m_1}\right)^{\frac{d}{2}}\frac{\Gamma \left(\frac{4-4 d+\Delta_1+\Delta_2+\Delta_3+\Delta_4}{2} \right) \Gamma \left(\frac{4-3 d+\Delta_1+\Delta_2+\Delta_3+\Delta_4}{2} \right)}{\Gamma (4-3 d+\Delta_1+\Delta_2+\Delta_3+\Delta_4)}\\
\times {}_2F_1\left(\frac{4-3d+\sum_i\Delta_i}{2},\frac{4-2d+\sum_i\Delta_i}{2};4-3d+\sum_i\Delta_i;1-\frac{m_2^2}{m_1^2}\right).
\end{multline}
The final result for the Mellin amplitude reads  
\begin{multline}\label{MAcandy}
    M^{{\cal V}_{12 \varphi_1 \varphi_2}{\cal V}_{34 \varphi_1 \varphi_2}}_{\Delta_1 \Delta_2\Delta_3\Delta_4}\left(\delta_{ij}\right)=g_{12}g_{34} \pi^{d+2} \,\prod^4_{i=1}\frac{1}{4\pi^{\frac{d+2}{2}}}\Gamma\left(\tfrac{d}{2}-\Delta_i\right)  \\ \times \Gamma \left(\frac{4-3d+\sum_i \Delta_i}{2}\right) \left(\frac{m_1}{2}\right)^{-4+3d-\sum_i\Delta_i}\left(\frac{1}{4\pi^{\frac{d+2}{2}}}\right)^2\left(\frac{m_2}{2}\right)^{d}\\ \times  \frac{\Gamma \left(\frac{4-4 d+\Delta_1+\Delta_2+\Delta_3+\Delta_4}{2} \right) \Gamma \left(\frac{4-3 d+\Delta_1+\Delta_2+\Delta_3+\Delta_4}{2} \right)}{\Gamma (4-3 d+\Delta_1+\Delta_2+\Delta_3+\Delta_4)}\\
\times {}_2F_1\left(\frac{4-3d+\sum_i\Delta_i}{2},\frac{4-2d+\sum_i\Delta_i}{2};4-3d+\sum_i\Delta_i;1-\frac{m_2^2}{m_1^2}\right)\\\times \frac{1}{\Gamma\left(-\tfrac{\beta}{2}\right)}\int^{+i\infty}_{-i\infty}\frac{{\rm d}z}{2\pi i}\frac{\Gamma\left(\delta_{12}+z-\tfrac{\beta}{2}+\Delta_1+\Delta_2-d\right)\Gamma\left(\delta_{34}+z-\tfrac{\beta}{2}+\Delta_3+\Delta_4-d\right)}{\Gamma\left(\delta_{12}+z-\tfrac{\beta}{2}\right)\Gamma\left(\delta_{34}+z-\tfrac{\beta}{2}\right)}\\ \times \Gamma\left(z-\tfrac{\beta}{2}\right)\Gamma\left(-z\right)\left(-1\right)^z.
\end{multline}
Interestingly, this is proportional to the Mellin amplitude \eqref{MAexchdij} for the four-point exchange diagram, with proportionality constant encoding the dependence on the masses $m_1$ and $m_2$ running in the loop. I.e.
\begin{multline}\label{1loopm1m2}
  {\cal A}^{{\cal V}_{12 \varphi_1 \varphi_2}{\cal V}_{34 \varphi_1 \varphi_2}}_{\Delta_1\Delta_2\Delta_3\Delta_4}\left(Q_1,Q_2,Q_3,Q_4\right)= \frac{1}{4\pi^{\frac{d+2}{2}}}\left(\frac{m_2}{2}\right)^{d}\\ \times  \frac{\Gamma \left(\frac{4-4 d+\Delta_1+\Delta_2+\Delta_3+\Delta_4}{2} \right) \Gamma \left(\frac{4-3 d+\Delta_1+\Delta_2+\Delta_3+\Delta_4}{2} \right)}{\Gamma (4-3 d+\Delta_1+\Delta_2+\Delta_3+\Delta_4)}\\
\times {}_2F_1\left(\frac{4-3d+\sum_i\Delta_i}{2},\frac{4-2d+\sum_i\Delta_i}{2};4-3d+\sum_i\Delta_i;1-\frac{m_2^2}{m_1^2}\right)\\ \times {\cal A}^{{\cal V}_{12 \varphi_1}{\cal V}_{34 \varphi_1}}_{\Delta_1 \Delta_2\Delta_3\Delta_4}\left(Q_1,Q_2,Q_3,Q_4\right).
\end{multline}
For exchanged scalars of equal mass $m=m_1=m_2$, this is simply:
\begin{multline}\label{equalmcandy}
  {\cal A}^{{\cal V}_{12 \varphi \varphi}{\cal V}_{34 \varphi \varphi}}_{\Delta_1\Delta_2\Delta_3\Delta_4}\left(Q_1,Q_2,Q_3,Q_4\right)= \frac{1}{2}\frac{1}{4\pi^{\frac{d+2}{2}}}\left(\frac{m}{2}\right)^{d}\\ \times  \frac{\Gamma \left(\frac{4-4 d+\Delta_1+\Delta_2+\Delta_3+\Delta_4}{2} \right) \Gamma \left(\frac{4-3 d+\Delta_1+\Delta_2+\Delta_3+\Delta_4}{2} \right)}{\Gamma (4-3 d+\Delta_1+\Delta_2+\Delta_3+\Delta_4)}\\ \times {\cal A}^{{\cal V}_{12 \varphi}{\cal V}_{34 \varphi}}_{\Delta_1 \Delta_2\Delta_3\Delta_4}\left(Q_1,Q_2,Q_3,Q_4\right),
\end{multline}
where we also divided by the symmetry factor. 

In summary, we see that the Mellin amplitude for the one-loop candy diagram \eqref{MAcandy} has the same functional dependence on the ``Mandelstam invariants" $s_{12}$ and $s_{13}$ as the Mellin amplitude for the exchange diagram \eqref{MAexchdij}. In the next section we will see that this property extends to four-point loop diagrams generated by the vertices \eqref{nptv} for all $n$. In section \ref{sec::nonpertMA}, using the K\"all\'en-Lehmann spectral representation of two-point functions it will be proven to be general property of celestial Mellin amplitudes for four-point exchange diagrams in perturbation theory and beyond.

\paragraph{Higher loops.}

The above one-loop example immediately extends an arbitrary number $n$ of lines connecting the two internal points (see figure \ref{fig::higherloop}). This is given by
\begin{multline}
   {\cal A}^{{\cal V}_{12 \varphi_1 \ldots \varphi_n}{\cal V}_{34 \varphi_1 \ldots \varphi_n}}\left(Y_1,Y_2,Y_3,Y_4\right) = (-i g_{12})(-i g_{34}) \int {\rm d}^{d+2}X {\rm d}^{d+2}Y\, G^{(0)}_{T}\left(X,Y_1\right)G^{(0)}_{T}\left(X,Y_2\right) \\ \times \left(\,\prod^n_{i=1}G^{(m_i)}_{T}\left(X,Y\right)\right)G^{(0)}_{T}\left(Y,Y_3\right)G^{(0)}_{T}\left(Y,Y_4\right).
\end{multline}
and employing Schwinger parametrization we have
\begin{multline}
    {\cal A}^{{\cal V}_{12 \varphi_1 \ldots \varphi_n}{\cal V}_{34 \varphi_1 \ldots \varphi_n}}\left(Y_1,Y_2,Y_3,Y_4\right)
    = - g_{12}g_{34} \left(\frac{1}{4\pi^{\frac{d+2}{2}}}i^{-\frac{d}{2}}\right)^4\, \prod^n_{i=1}\left(\frac{1}{4\pi^{\frac{d+2}{2}}}\left(\frac{m_i}{2}\right)^{\frac{d}{2}}\right)\\ \times  \int^{+i\infty}_{-i\infty}\prod^n_{i=1}\frac{{\rm d}{\bar s}_i}{2\pi i}\Gamma\left({\bar s}_i-\tfrac{d}{4}\right)\left(\frac{m_i}{2}\right)^{-2{\bar s}_i} i^{-({\bar s}_i+\frac{d}{4})} \\ \times \int^{\infty}_0 \prod^n_{i=1}\frac{{\rm d}{\bar t}_i}{{\bar t}_i} {{\bar t}_i}^{{\bar s}_i+\frac{d}{4}}  \prod^4_{i=1} \frac{{\rm d}t_i}{t_i} t^{\frac{d}{2}}_i\, \int {\rm d}^{d+2}X {\rm d}^{d+2}Y\, \exp\left[i t_1\left(X-Y_1\right)^2+i t_2\left(X-Y_2\right)^2 \right. \\ \left.+ i ({\bar t}_1+{\bar t}_2+\ldots+{\bar t}_n )\left(X-Y\right)^2+i t_3\left(Y-Y_3\right)^2+i t_4\left(Y-Y_4\right)^2\right].
\end{multline}
As for the one-loop example above ($n=2$), the Gaussian integral is the same as that for the exchange diagram \eqref{GaussExc} with the Schwinger parameter for the internal leg replaced by the sum of the Schwinger parameters ${\bar t}_i$ for the internal legs connecting the two internal points.

\vskip 4pt 
The Mellin amplitude for the corresponding contribution to the celestial four-point function can therefore be immediately written down as
\begin{multline}
    M^{{\cal V}_{12 \varphi_1 \ldots \varphi_n}{\cal V}_{34 \varphi_1 \ldots \varphi_n}}_{\Delta_1 \Delta_2\Delta_3\Delta_4}\left(\delta_{ij}\right)=g_{12}g_{34}  \pi^{d+2}\,\prod^4_{i=1}\frac{1}{4\pi^{\frac{d+2}{2}}}\Gamma\left(\tfrac{d}{2}-\Delta_i\right) \prod^n_{i=1}\left(\frac{1}{4\pi^{\frac{d+2}{2}}}\left(\frac{m_i}{2}\right)^{\frac{d}{2}}\right) \\ \times \frac{1}{\Gamma\left(2-d+\tfrac{1}{2}\sum_i\Delta_i\right)} \int^{+i\infty}_{-i\infty}\prod^n_{i=1}\frac{{\rm d}{\bar s}_i}{2\pi i}\Gamma\left({\bar s}_i+\tfrac{d}{4}\right)\Gamma\left({\bar s}_i-\tfrac{d}{4}\right)\left(\frac{m_i}{2}\right)^{-2{\bar s}_i} \\ \times  2\pi i\, \delta\left(2-d+\tfrac{1}{2}\sum_i \Delta_i-\sum_i \left({\bar s}_i+\tfrac{d}{4}\right)\right) \\ \times \frac{1}{\Gamma\left(-\tfrac{\beta}{2}\right)}\int^{+i\infty}_{-i\infty}\frac{{\rm d}z}{2\pi i}\frac{\Gamma\left(\delta_{34}+z-\tfrac{\beta}{2}+\Delta_1+\Delta_2-d\right)\Gamma\left(\delta_{12}+z-\tfrac{\beta}{2}+\Delta_3+\Delta_4-d\right)}{\Gamma\left(\delta_{12}+z-\tfrac{\beta}{2}\right)\Gamma\left(\delta_{34}+z-\tfrac{\beta}{2}\right)}\\ \times \Gamma\left(z-\tfrac{\beta}{2}\right)\Gamma\left(-z\right)\left(-1\right)^z,
\end{multline}
which, like for the one-loop example above, is proportional to the Mellin amplitude for the four-point exchange diagram \eqref{MAexchdij} with proportionality constant encoding the masses $m_i$ of the exchanged particles. I.e.
\begin{multline}\label{nintfrom1int}
    {\cal A}^{{\cal V}_{12 \varphi_1 \ldots \varphi_n}{\cal V}_{34 \varphi_1 \ldots \varphi_n}}_{\Delta_1\Delta_2\Delta_3\Delta_4}\left(Q_1,Q_2,Q_3,Q_4\right) = f_{\varphi_1|\varphi_2 \ldots \varphi_n}\left(m_1, \ldots,m_n\right)\\ \times {\cal A}^{{\cal V}_{12 \varphi_1}{\cal V}_{34 \varphi_1}}_{\Delta_1 \Delta_2\Delta_3\Delta_4}\left(Q_1,Q_2,Q_3,Q_4\right),
\end{multline}
where 
\begin{multline}\label{f1}
   f_{\varphi_1|\varphi_2 \ldots \varphi_n}\left(m_1, \ldots,m_n\right)= 4\pi^{\frac{d+2}{2}}\left(\frac{m_1}{2}\right)^{-3d+4+\sum_i\Delta_i}    \\ \times \frac{1}{\Gamma\left(\tfrac{4-2d+\sum_i\Delta_i}{2}\right)\Gamma\left(\tfrac{-3d+4+\sum_i\Delta_i}{2}\right)} \int^{+i\infty}_{-i\infty}\prod^n_{i=1}\frac{{\rm d}{\bar s}_i}{2\pi i}\frac{1}{4\pi^{\frac{d+2}{2}}}\Gamma\left({\bar s}_i+\tfrac{d}{4}\right)\Gamma\left({\bar s}_i-\tfrac{d}{4}\right)\left(\frac{m_i}{2}\right)^{-2{\bar s}_i+\frac{d}{2}} \\ \times  2\pi i\, \delta\left(2-d+\tfrac{1}{2}\sum_i \Delta_i-\sum_i \left({\bar s}_i+\tfrac{d}{4}\right)\right).
\end{multline}
The celestial Mellin amplitude for a process involving the exchange of $n$ particles between a pair of internal points therefore has the same poles in $s_{12}$ as that for a single particle exchanged between the same pair of internal points (summarised in figure \ref{fig::exch_poles}). This is to be contrasted with the corresponding story in anti-de Sitter space where, owing to the discrete spectrum, the Mellin amplitudes for such processes have different poles corresponding to the scaling dimensions of the bound states in the bulk (see e.g. \cite{Meltzer:2019nbs}).

\paragraph{Massless exchanged scalars.} As for the exchange of a single scalar field in section \ref{subsec::exch}, it is straightforward to consider the case that one or more of the exchanged scalars are massless. This can be achieved by a massless limit or working directly with massless Feynman propagators. The former is straightforward to implement at the level of the function \eqref{f1} via
\begin{equation}
\lim_{m_i \to 0}\Gamma\left({\bar s}_i+\tfrac{d}{4}\right)\Gamma\left({\bar s}_i-\tfrac{d}{4}\right)\left(\frac{m_i}{2}\right)^{-2{\bar s}_i+\frac{d}{2}} = \Gamma\left(\tfrac{d}{2}\right)2\pi i \delta\left({\bar s}_i-\tfrac{d}{4}\right).
\end{equation}
So that 
\begin{multline}
\lim_{m_{j\ne 1} \to 0}   f_{\varphi_1|\varphi_2 \ldots \varphi_n}\left(m_1, \ldots,m_n\right)= 4\pi^{\frac{d+2}{2}}\left(\frac{m_1}{2}\right)^{-3d+4+\sum_i\Delta_i}    \\ \times \frac{\Gamma\left(\tfrac{d}{2}\right)}{\Gamma\left(\tfrac{4-2d+\sum_i\Delta_i}{2}\right)\Gamma\left(\tfrac{-3d+4+\sum_i\Delta_i}{2}\right)} \int^{+i\infty}_{-i\infty}\prod^n_{i\ne j}\frac{{\rm d}{\bar s}_i}{2\pi i}\frac{1}{4\pi^{\frac{d+2}{2}}}\Gamma\left({\bar s}_i+\tfrac{d}{4}\right)\Gamma\left({\bar s}_i-\tfrac{d}{4}\right)\left(\frac{m_i}{2}\right)^{-2{\bar s}_i+\frac{d}{2}} \\ \times  2\pi i\, \delta\left(\frac{4-3d+\sum_i \Delta_i-2\sum_{i \ne j} \left({\bar s}_i+\tfrac{d}{4}\right)}{2}\right).
\end{multline}
In the case that all exchanged scalars are massless, iterating the above formula gives the Mellin amplitude:
\begin{multline}
    M^{{\cal V}_{12 \varphi_1 \ldots \varphi_n}{\cal V}_{34 \varphi_1 \ldots \varphi_n}}_{\Delta_1 \Delta_2\Delta_3\Delta_4}\left(s_{12},s_{13}\right)=g_{12}g_{34}\,\pi^{d+2}\,\frac{1}{\Gamma\left(\frac{nd}{2}\right)}\left(\frac{\Gamma\left(\frac{d}{2}\right)}{4 \pi^{\frac{d+2}{2}}}\right)^n\prod^4_{i=1}\frac{1}{4\pi^{\frac{d+2}{2}}}\Gamma\left(\tfrac{d}{2}-\Delta_i\right)\,\\ \times 2\pi i\, \delta\left(\frac{4-(2+n)d+\sum_i \Delta_i}{2}\right) \frac{\Gamma\left(\tfrac{2-d+\Delta_1+\Delta_2-s_{12}}{2}\right)\Gamma\left(\tfrac{2-d+\Delta_3+\Delta_4-s_{12}}{2}\right)}{\Gamma\left(\tfrac{d+2-\Delta_1-\Delta_2-s_{12}}{2}\right)\Gamma\left(\tfrac{d+2-\Delta_3-\Delta_4-s_{12}}{2}\right)}\\ \times {}_3F_2\left(\begin{matrix}\tfrac{2-d+\Delta_1+\Delta_2-s_{12}}{2},\tfrac{2-d+\Delta_3+\Delta_4-s_{12}}{2},-\frac{\beta}{2}\\\tfrac{d+2-\Delta_1-\Delta_2-s_{12}}{2},\tfrac{d+2-\Delta_3-\Delta_4-s_{12}}{2} \end{matrix};1\right),
\end{multline}
where we see that the linear constraint on the scaling dimensions $\Delta_i$ depends on the number $n$ of lines connecting the two bulk points. For $n\ne1$ Saalsch\"utz's theorem \eqref{ssth} no longer holds.

\section{Non-perturbative celestial Mellin amplitudes}
\label{sec::nonpertMA}

\begin{figure}[t]
    \centering
    \includegraphics[width=0.3\textwidth]{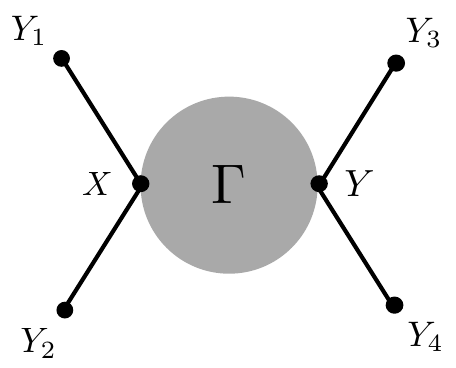}
    \caption{Four-point exchange diagram with the exact bulk two-point function $\Gamma\left(X,Y\right)$.}
    \label{fig::4ptexch}
\end{figure}

The non-perturbative properties of celestial Mellin amplitudes can be studied using  the K\"all\'en-Lehmann spectral representation for exact bulk two-point functions:
\begin{align}\nonumber 
\Gamma(X,Y)&=\langle \Omega | T\left\{\phi(X)\phi(Y)\right\}|\Omega \rangle,\\
&= \int^\infty_0 {\rm d}\mu^2\, \rho(\mu^2) \,G^{(\mu)}_T(X,Y), \label{KLMink}
\end{align}
where unitarity implies a positive definite spectral function $\rho(\mu^2) \geq 0$. As noted in \cite{Iacobacci:2024nhw}, these properties are encoded in celestial correlators via the Mellin transform of the spectral density, which can be seen by extrapolating the exact bulk two-point function \eqref{KLMink} to the celestial sphere according to the prescription \eqref{ccdefn}. Applying the convolution theorem for the Mellin transform this is:
\begin{align}
    \Gamma_{\Delta_1,\Delta_2}(Q_1,Q_2)
    &= \int^\infty_0 {\rm d}\mu^2\, \rho(\mu^2) \,G^{(\mu)}_{\Delta_1 \Delta_2}(Q_1,Q_2),\\
    &={\cal M}\left[\rho\right]\left(\tfrac{d}{2}-\Delta_1+1\right) \left(\frac{1}{2}\right)^{d-2\Delta_1}\frac{C^{\text{flat}}_{\Delta_1}}{(-2Q_1\cdot Q_2+i\epsilon)^{\Delta_1}}(2\pi )\delta(i(\Delta_1-\Delta_2)), \nonumber 
\end{align}
in terms of the Mellin transformed spectral function:
\begin{equation}\label{mellinrho}
    {\cal M}\left[\rho\right]\left(s\right)=\int^\infty_0 \frac{{\rm d}z}{z}\, \rho(z)z^s.
\end{equation}
The free theory boundary two-point function is given by \cite{Sleight:2023ojm}
\begin{align}
     G^{(\mu)}_{\Delta_1 \Delta_2}(Q_1,Q_2)
     &= \left(\frac{\mu}2\right)^{d-2\Delta} \frac{C^{\text{flat}}_{\Delta_1}}{(-2Q_1\cdot Q_2+i\epsilon)^{\Delta_1}}(2\pi )\delta(i(\Delta_1-\Delta_2))\,, \nonumber\\
     \label{celest2ptnorm}
   C^{\text{flat}}_{\Delta}&= \frac{1}{4\pi^{\frac{d+2}2}}\,\Gamma(\Delta)\Gamma(\Delta-\tfrac{d}2).
\end{align}
The analytic properties of the Mellin transformed spectral function \eqref{mellinrho} were discussed in \cite{Iacobacci:2024nhw}.

\vskip 4pt
Being a superposition of free theory Feynman propagators \eqref{KLMink}, using the K\"all\'en-Lehman spectral representation we can apply the techniques presented in the previous sections to study non-perturbative celestial Mellin amplitudes. For example, using the K\"all\'en-Lehman spectral representation the four-point exchange diagram mediated by the exact two-point function \eqref{KLMink} reads  (see figure \ref{fig::4ptexch})
\begin{equation}
   {\cal A}^{{\cal V}_{12 \Gamma}{\cal V}_{34 \Gamma}}\left(Y_1,Y_2,Y_3,Y_4\right) = \int^\infty_0 {\rm d}\mu^2\, \rho(\mu^2) {\cal A}^{{\cal V}_{12 \varphi}{\cal V}_{34 \varphi}}_{\mu}\left(Y_1,Y_2,Y_3,Y_4\right),
\end{equation}
and the corresponding celestial Mellin amplitude is given by:
\begin{shaded}
 \begin{multline}\label{np4ptexch}
    M^{{\cal V}_{12 \Gamma}{\cal V}_{34 \Gamma}}_{\Delta_1 \Delta_2\Delta_3\Delta_4}\left(s_{12},s_{13}\right)=g_{12}g_{34}\,\pi^{d+2}\,\prod^4_{i=1}\frac{1}{4\pi^{\frac{d+2}{2}}}\Gamma\left(\tfrac{d}{2}-\Delta_i\right)\,\\ \times {\cal M}\left[\rho\right]\left(\tfrac{3d-2-\sum_i\Delta_i}{2}\right) \,\frac{1}{4\pi^{\frac{d+2}{2}}} \left(\frac{1}{2}\right)^{3d-4-\sum_i\Delta_i}\Gamma\left(\frac{-3d+4+\sum_i\Delta_i}{2}\right)\\ \times \frac{\Gamma\left(\tfrac{2-d+\Delta_1+\Delta_2-s_{12}}{2}\right)\Gamma\left(\tfrac{2-d+\Delta_3+\Delta_4-s_{12}}{2}\right)}{\Gamma\left(\tfrac{d+2-\Delta_1-\Delta_2-s_{12}}{2}\right)\Gamma\left(\tfrac{d+2-\Delta_3-\Delta_4-s_{12}}{2}\right)}\\ \times {}_3F_2\left(\begin{matrix}\tfrac{2-d+\Delta_1+\Delta_2-s_{12}}{2},\tfrac{2-d+\Delta_3+\Delta_4-s_{12}}{2},-\frac{\beta}{2}\\\tfrac{d+2-\Delta_1-\Delta_2-s_{12}}{2},\tfrac{d+2-\Delta_3-\Delta_4-s_{12}}{2} \end{matrix};1\right).
\end{multline}   
\end{shaded}
\noindent This has the same functional dependence on the ``Mandelstam invariants" $s_{12}$ and $s_{13}$ as the Mellin amplitude \eqref{Mexchs12s13} for the tree-level exchange, as observed in the examples of the previous sections. The non-perturbative structure is encoded in the Mellin transform \eqref{mellinrho} of the spectral function.

\vskip 4pt
As a consistency check, the spectral function for the one-loop bubble diagram of equal mass scalars is
\begin{align}\label{osfbub}
    \rho(z)=\frac{1}{2}\frac{1}{2^{2d+1}\pi^{\frac{d+1}2}} \frac{\left(z -4 m^2\right)^{\frac{d-1}{2}}}{\sqrt{z}}\ \theta \left(\frac{z }{4 m^2}-1\right)\,,
\end{align}
which has Mellin transform \cite{Iacobacci:2024nhw}:
\begin{align}
    \mathcal{M}[\rho](s)=\frac{(2 m)^{d+2 s-2}}{2^{2d+1} \pi ^{\frac{d+1}{2}}}\frac{\Gamma \left(\frac{d+1}{2}\right) \Gamma \left(1-\frac{d}{2}-s\right)}{\Gamma \left(\frac{3}{2}-s\right)}\,.
\end{align}
Plugging into \eqref{np4ptexch} reproduces the expression \eqref{equalmcandy} for the celestial four-point candy diagram.

\section*{Acknowledgements} 

We thank Sébastien Malherbe, Michel Pannier and Massimo Taronna for discussions. This research was partially supported by the INFN initiative STEFI.

\newpage

\begin{appendix}

\section{Integrals}
\label{A::DI}

In this appendix we give further details on the evaluation of various integrals encountered in the perturbative computation of Celestial Mellin amplitudes.

\subsection{Minkowski integrals}
\label{A::GI}

Using Schwinger parametrization \eqref{SPXY} of Feynman propagators, integrals over Minkowski space in the perturbative computation of time-ordered correlators are reduced to Gaussian integrals of the form:
\begin{align}\label{Gaussb}
    \int {\rm d}^{d+2}X \exp\left[i \sum\limits_i t_i\left(X-Y_i\right)^2\right]&=\exp\left[-i\frac{(t_1Y_1+\dots+t_nY_n)^2}{t_1+\dots+t_n}+i\sum_i t_iY_i^2\right]\\ \nonumber 
    & \hspace*{-0.5cm}\times\underbrace{\int {\rm d}^{d+2}X \exp\left[i(t_1+\dots+t_n)\left(X-\frac{t_1Y_1+\dots+t_nY_n}{t_1+\dots+t_n}\right)^2\right]}_{i^{\frac{d}{2}}\pi^{\frac{d+2}{2}}\left(t_1 + \ldots +t_n
    \right)^{-\frac{d+2}{2}}},
\end{align}
where we used that
\begin{equation}
    \int^{+\infty}_{-\infty}{\rm d}x\,\exp\left[-i a x^2\right] = \left(\frac{\pi}{ia}\right)^{\frac{1}{2}}.
\end{equation}

\vskip 4pt
Minkowski integrals for Feynman diagrams with two or more internal points are given by nested Gaussian integrals.  For diagrams with two internal points, such nested integrals all take the form 
\begin{multline}
  \int {\rm d}^{d+2}X {\rm d}^{d+2}Y\, \exp\left[i t\left(X-Y\right)^2+it_1\left(X-Y_1\right)^2+i t_2\left(X-Y_2\right)^2\right.\\\left.\hspace*{10cm}+i t_3\left(Y-Y_3\right)^2+i t_4\left(Y-Y_4\right)^2\right]\\
  = i^{d} \pi^{d+2} \left[({\bar t}+t_1+t_2)({\bar t}+t_3+t_4)-t^2\right]^{-\frac{d+2}{2}} \exp\left[i\sum^4_{i=1}t_i Y^2_i\right.\\  \nonumber \left.-i\frac{({\bar t}+t_1+t_2)\left(t_3 Y_3+t_4 Y_4\right)^2+2{\bar t} \left(t_1 Y_1+t_2 Y_2\right)\left(t_3 Y_3+t_4 Y_4\right)+({\bar t}+t_3+t_4)\left(t_1 Y_1+t_2 Y_2\right)^2}{({\bar t}+t_1+t_2)({\bar t}+t_3+t_4)-{\bar t}^2}\right],
\end{multline}
which comes from repeated application of the formula \eqref{Gaussb}.

\subsection{Integrals over Schwinger parameters}
\label{A::SI}

The approach outlined at the beginning of section \ref{sec::MA4pcc} reduces the computation of perturbative Celestial Mellin amplitudes to the evaluation of integrals over Schwinger parameters of the corresponding Feynman propagators. In the following we give such integrals for the Celestial Mellin amplitudes considered in this work, which involve at most two internal points.

\paragraph{Contact diagrams.} For contact diagrams, the integrals over the Schwinger parameters take the following form (for some $a_i$ and $b$):
\begin{align}
     T_{\text{cont.}}\left(a_1,\ldots a_n;b\right)&=\int^{\infty}_0 \prod^n_{i=1} \frac{{\rm d}t_i}{t_i} t^{a_i}_i \left(t_1+\ldots + t_n\right)^{b}, \\ &= 2 \pi i\, \delta (b+\sum_i a_i) \frac{1}{\Gamma\left(-b\right)} \prod^n_{i=1}\Gamma\left(a_i\right). 
\end{align} 
 Writing
\begin{equation}
    \left(t_1+\ldots + t_n\right)^{b}=\frac{1}{\Gamma\left(-b\right)}\int^\infty_0\frac{{\rm d}u}{u}u^{-b}e^{-u(t_1+\ldots + t_n)},
\end{equation}
and applying the change of variables $t_i\to t_i/u$, the integrals completely factorise into $\Gamma$-functions and a Dirac delta function:
\begin{equation}
     T_{\text{cont.}}\left(a_1,\ldots a_n;b\right)= \frac{1}{\Gamma\left(-b\right)} \underbrace{\int^{\infty}_0 \prod^n_{i=1} \frac{{\rm d}t_i}{t_i} t^{a_i}_i e^{-t_i}}_{\prod^n_{i=1}\Gamma\left(a_i\right)} \times \underbrace{\int^\infty_0 \frac{{\rm d}u}{u} u^{-b-\sum_i a_i}}_{2\pi i\,\delta\left(-b-\sum_i a_i\right)}.
 \end{equation}

 \paragraph{Particle exchanges.} For four-point processes involving the exchange of one or more particles between two internal points (such as those in sections \ref{subsec::exch} and \ref{subsec::1loop}), the integrals over the Schwinger parameters take the form
\begin{multline}
T_{\text{exch}}\left(a_1,a_2,a_3,a_4;a;b;b_{12},b_{34}\right)=\int^\infty_0 \frac{{\rm d}{\bar t}}{{\bar t}} {\bar t}^{-(\delta_{13}+\delta_{14}+\delta_{23}+\delta_{24})+a}\\ \times \int^\infty_0 \prod^4_{i=1}\frac{{\rm d}t_i}{t_i} t^{a_i}_i\left[({\bar t}+t_1+t_2)({\bar t}+t_3+t_4)-{\bar t}^2\right]^{b}\\ \times \left({\bar t}+t_1+t_2\right)^{-\delta_{34}+b_{12}}\left({\bar t}+t_3+t_4\right)^{-\delta_{12}+b_{34}}.
\end{multline}
To proceed we use the following generalisation of the Binomial expansion:
\begin{equation}
    \left(x+y\right)^{b} = \frac{1}{\Gamma\left(-b\right)} \int^{+i\infty}_{-i\infty}\frac{{\rm d}z}{2\pi i} \Gamma\left(z-b\right)\Gamma\left(-z\right)x^z y^{b-z},
\end{equation}
with
\begin{equation}
    x = -{\bar t}^2, \qquad y = ({\bar t}+t_1+t_2)({\bar t}+t_3+t_4).
\end{equation}
The integrals over $t_{i}$ associated to the external fields then factorise: 
\begin{multline}
T_{\text{exch}}\left(a_1,a_2,a_3,a_4;a;b;b_{12},b_{34}\right)=\frac{1}{\Gamma\left(-b\right)} \int^{+i\infty}_{-i\infty}\frac{{\rm d}z}{2\pi i}\Gamma\left(z-b\right)\Gamma\left(-z\right)\left(-1\right)^z\\ \times \int^{\infty}_0 \frac{{\rm d}{\bar t}}{{\bar t}} {\bar t}^{-\left(\delta_{13}+\delta_{14}+\delta_{23}+\delta_{24}\right)+a+2z}\\ \times  \int^\infty_0 \frac{{\rm d}t_1}{t_1}\frac{{\rm d}t_2}{t_2} t^{a_1}_1t^{a_2}_2\left({\bar t}+t_1+t_2\right)^{b+b_{12}-\left(\delta_{34}+z\right)}\int^\infty_0 \frac{{\rm d}t_3}{t_3}\frac{{\rm d}t_4}{t_4} t^{a_3}_3t^{a_4}_4\left({\bar t}+t_3+t_4\right)^{b+b_{34}-\left(\delta_{12}+z\right)}.
\end{multline}
The integrals over the external Schwinger parameters $t_i$ take the same form as for the contact diagrams above. Writing:
\begin{equation}
    ({\bar t}+t_i+t_j)^{c}=\frac{1}{\Gamma\left(-c\right)}\int^\infty_0\frac{{\rm d}u}{u}u^{-c}e^{-u({\bar t}+t_i+t_j)},
\end{equation}
and performing the change of variables $t_{i,j}\to t_{i,j}/u$ and $u \to u/{\bar t}$, the integrals completely factorise into $\Gamma$-functions:
\begin{align}\nonumber 
  \int^\infty_0 \frac{{\rm d}t_i}{t_i}\frac{{\rm d}t_j}{t_j} t^{a_i}_it^{a_j}_j ({\bar t}+t_i+t_j)^{c} &= \frac{{\bar t}^{a_i+a_j+c}}{\Gamma\left(-c\right)}\int^\infty_0\frac{{\rm d}u}{u}u^{-a_i-a_j-c}e^{-u} \int^\infty_0 \frac{{\rm d}t_i}{t_i}t^{a_i}_ie^{-t_i}\int^\infty_0\frac{{\rm d}t_j}{t_j} t^{a_j}_j e^{-t_j}\\ 
   &={\bar t}^{a_i+a_j+c} \frac{\Gamma\left(
    a_i\right)\Gamma\left(
    a_j\right)\Gamma\left(-a_i-a_j-c\right)}{\Gamma\left(-c\right)}.
\end{align}

The integral over the Schwinger parameter ${\bar t}$ (associated to the exchanged field) then reduces to a Dirac delta function:
\begin{multline}
    \int^\infty_0 \frac{{\rm d}{\bar t}}{{\bar t}} {\bar t}^{-\tfrac{d+2}{2}-\tfrac{\beta}{2}+a+a_1+a_2+a_3+a_4+2b+b_{12}+b_{34}} \\ = 2 \pi i\, \delta\left(-\tfrac{d+2}{2}-\tfrac{\beta}{2}+a+a_1+a_2+a_3+a_4+2b+b_{12}+b_{34}\right).
\end{multline}

 Altogether this gives 
 \begin{multline}
T_{\text{exch}}\left(a_1,a_2,a_3,a_4;a;b;b_{12},b_{34}\right)=2 \pi i\, \delta\left(-\tfrac{d+2}{2}-\tfrac{\beta}{2}+a+a_1+a_2+a_3+a_4+2b+b_{12}+b_{34}\right)\\ \times \frac{\left(\prod^4_{i=1}\Gamma\left(a_i\right)\right)}{\Gamma\left(-b\right)} \int^{+i\infty}_{-i\infty}\frac{{\rm d}z}{2\pi i} \frac{\Gamma\left(z+\delta_{34}-b-b_{12}-a_1-a_2+z\right)\Gamma\left(z+\delta_{12}-b-b_{34}-a_3-a_4+z\right)}{\Gamma\left(z+\delta_{34}-b-b_{12}\right)\Gamma\left(z+\delta_{12}-b-b_{34}\right)}\\ \times \Gamma\left(z-b\right)\Gamma\left(-z\right)\left(-1\right)^z.
\end{multline}

\section{Massive External Fields}
\label{A::MassiveExt}

While in sections \ref{subsec::exch} and \ref{subsec::1loop} for ease of presentation we focused on diagrams with external massless scalars, the approach straightforwardly extends to external massive fields. As for the contact diagrams in section \ref{subsec::contact}, each external massive fields is accompanied by a Mellin-Barnes integral \eqref{SPm} which encodes the mass dependence. 

\vskip 4pt
For example, the Mellin amplitude for the exchange diagram generated by cubic vertices \eqref{cubicv} but with massive external scalar fields $\phi_i$ of mass $m_i$ is simply:
\begin{multline}\label{extmass}
    M^{{\cal V}_{12 \phi}{\cal V}_{34 \phi}}_{\Delta_1 \Delta_2\Delta_3\Delta_4}\left(s_{12},s_{13}\right)=g_{12}g_{34}\,\pi^{d+2}\,\\ 
    \times  \int^{+i\infty}_{-i\infty}\frac{{\rm d}s_i}{2\pi i}\prod^4_{i=1}\frac{1}{4\pi^{\frac{d+2}{2}}}\Gamma\left(s_i+\tfrac{1}{2}\left(\Delta_i-\tfrac{d}{2}\right)\right)\Gamma\left(s_i-\tfrac{1}{2}\left(\Delta_i-\tfrac{d}{2}\right)\right)\left(\frac{m_i}{2}\right)^{-2s_i+\frac{d}{2}-\Delta_i} i^{-(s_i+\frac{d}{4}+\frac{\Delta_i}{2})}\\
    \times \int^{+i\infty}_{-i\infty}\frac{{\rm d}{\bar s}}{2\pi i}\Gamma\left({\bar s}-\tfrac{d}{4}\right)\left(\frac{m}{2}\right)^{-2{\bar s}+\frac{d}{2}} i^{-({\bar s}+\frac{d}{4})}\\
   \times  2\pi i\, \delta\left({\bar s}+\tfrac{d}{4}-2+\sum_{i}s_i\right)
    \\
    \times \frac{1}{\Gamma\left(-\frac{\beta}{2}\right)}\int^{+i\infty}_{-i\infty}\frac{{\rm d}z}{2\pi i}\frac{\Gamma\left(z+\tfrac{1}{2}\left(2-2s_1-2s_2-s_{12}\right)\right)\Gamma\left(z+\tfrac{1}{2}\left(2-2s_3-2s_4-s_{12}\right)\right)}{\Gamma\left(z+\tfrac{1}{2}(d+2-\Delta_3-\Delta_4-s_{12})\right)\Gamma\left(z+\tfrac{1}{2}(d+2-\Delta_1-\Delta_2-s_{12})\right)}\\ \times \Gamma\left(z-\tfrac{\beta}{2}\right)\Gamma\left(-z\right)\left(-1\right)^z.
\end{multline}
 Similarly one can write down the expression for the loop diagrams in section \ref{subsec::1loop} with external massive fields. In the massless limit $m_i\to 0$, using \eqref{mllim}, this reproduces the expression \eqref{MAexchdij} for external massless scalars.

\vskip 4pt
Pole pinching of the $z$ integral in the expression \eqref{extmass} gives rise to the following poles in $s_{12}$ where $n=0,\,1,\,2,\, \ldots$:
\begin{align}
    s_{12}&=2-2s_1-2s_2+2n, \qquad s_{12}=2-2s_3-2s_4+2n.
\end{align}
The Mellin amplitude can be evaluated as a power series in $\left(m_i/m\right)$ by closing the integration contour for the $s_i$ integrals to the right, on the poles:
\begin{equation}
    s_i = \pm \tfrac{1}{2}\left(\Delta_i-\tfrac{d}{2}\right)-n_i,\, \qquad n_i = 0,\, 1,\,2,\, \ldots\,,
\end{equation}
where $n_i$ parametrizes the power of the mass $m_i$. These correspond to poles in $s_{12}$ at:
\begin{subequations}
  \begin{align}\label{mlcon1}
    s_{12}&=2-d+\Delta_1+\Delta_2+2\left(n+n_1+n_2\right),\\
    s_{12}&=2+d-\Delta_1-\Delta_2+2\left(n+n_1+n_2\right),\\
    s_{12}&=2-\left(\Delta_1-\Delta_2\right)+2\left(n+n_1+n_2\right),\\
    s_{12}&=2+\left(\Delta_1-\Delta_2\right)+2\left(n+n_1+n_2\right),
\end{align}  
\end{subequations}
and
\begin{subequations}
 \begin{align}\label{mlcon2}
    s_{12}&=2-d+\Delta_3+\Delta_4+2\left(n+n_3+n_4\right),\\
    s_{12}&=2+d-\Delta_3-\Delta_4+2\left(n+n_3+n_4\right),\\
    s_{12}&=2-\left(\Delta_3-\Delta_4\right)+2\left(n+n_3+n_4\right),\\
    s_{12}&=2+\left(\Delta_3-\Delta_4\right)+2\left(n+n_3+n_4\right).
\end{align}   
\end{subequations}
In the limit $m_i \to 0$ only the contributions from the poles \eqref{mlcon1} and \eqref{mlcon2} with $n_i=0$ survive, reproducing the contributions \eqref{ppexch} for massless external scalars.

\end{appendix}

\bibliographystyle{JHEP}
\bibliography{refs}

\end{document}